\documentclass[11pt]{article}

\usepackage{fullpage}
\usepackage{amsmath}
\usepackage{amssymb}
\usepackage{mathrsfs}
\usepackage{setspace}
\usepackage{bbm}
\usepackage{dsfont}

\usepackage{graphics}

\usepackage[latin1]{inputenc}
\usepackage[pdftex]{graphicx}

\usepackage{color}
\usepackage[american]{babel}

\newcommand{\be}{\begin{equation}}
\newcommand{\ee}{\end{equation}}
\newcommand{\bea}{\begin{eqnarray}}
\newcommand{\eea}{\end{eqnarray}}
\newcommand{\bb}{\mathbb}
\newcommand{\p}{\partial}
\newcommand{\irr}{{\textit{irr}}}
\newcommand{\sgn}{{\rm sign}}
\newcommand{\mat}{\left(\begin{array}{cc}}
\newcommand{\emat}{\end{array}\right)}
\newcommand{\smat}{\left(\begin{smallmatrix}}
\newcommand{\esmat}{\end{smallmatrix}\right)}
\newcommand{\qnm}{\textsc{qnm}}

\newcommand{\Z}{\mathbb{Z}}

\begin{document}
\numberwithin{equation}{section}
{
\begin{titlepage}
\begin{center}

\hfill \\
\hfill \\
\vskip 0.75in

{\Large \bf Black Hole Scattering from Monodromy}\\

\vskip 0.4in

{Alejandra Castro,${}^a$ Joshua M. Lapan,${}^b$ Alexander Maloney,${}^b$ and Maria J. Rodriguez${}^a$}\\

\vskip 0.3in

{\it ${}^a$ Center for the Fundamental Laws of Nature, Harvard University, Cambridge, MA 02138, USA }  \vskip .5mm              
{\it ${}^b$ Physics Department, McGill University, 3600 rue University, Montr\'eal, QC H3A 2T8, Canada }

\vskip 0.3in

\end{center}

\vskip 0.35in

\begin{abstract} 
\noindent
We study scattering coefficients in black hole spacetimes using analytic properties of complexified wave equations.  For a concrete example, we analyze the singularities of the Teukolsky equation and relate the corresponding monodromies to scattering data.  These techniques, valid in full generality, provide insights into complex-analytic properties of greybody factors and quasinormal modes.  This leads to new perturbative and numerical methods which are in good agreement with previous results.
\end{abstract}

\vfill

\noindent April 12, 2013

\end{titlepage}
}

\newpage
\tableofcontents


\section{Introduction}

In the study of quantum field theory in curved spacetime, perhaps the most basic object worth considering is the two-point function in a black hole background.  This quantity describes the amplitude for a particle to scatter off of the black hole and contains a wealth of information of formal and observational interest.  In particular, it appears in the formula for Hawking radiation, thus playing a role in the description of quantum mechanical black holes; indeed, the low frequency behavior of scattering coefficients can be matched precisely with proposed conformal field theory descriptions of black hole microstates.  Observationally, the poles in the scattering coefficients are quasinormal modes, which characterize the late-time relaxation towards a black hole equilibrium state.  
Formal properties of these quasinormal modes are related to the stability of black hole spacetimes, which is in many cases an open and difficult problem.  

In computing these scattering amplitudes the primary obstacle is  technical---the field equations cannot be solved analytically.  In this paper we describe techniques, based on global analytic properties of probes of 
black hole geometries, that allow us to extract certain precise features of scattering amplitudes even without a direct analytic solution.  This approach has the advantage of distinguishing global versus local data, allowing us to parameterize the dependence of certain physical observables on analytic properties of the solution. 

Some of the results we present here are not new; we refer the reader to the reviews of \cite{Nollert:1999ji,Berti:2009kk,Konoplya:2011qq} and citations therein. 
Global aspects of the relevant class of ODEs have been studied to some extent in prior literature. See e.g. \cite{leaver:1238,Fiziev:2005ki,Fiziev2010} for an extensive analysis of  properties of the solutions to confluent Heun equations and \cite{Mano:1996vt,Mano:1996gn} for an interesting  related derivation of greybody factors for the Kerr black hole.

Our analysis starts with a remarkable fact: in a broad class of black hole backgrounds, including Kerr,  the field equation separates; radial dependence is determined by an ordinary differential equation with regular singular points at the event horizons and an irregular singular point at asymptotic infinity.   This means that if we complexify the radial coordinate, the field is locally a holomorphic function of the radius with branch points at the horizons and infinity.  These branch points reflect the fact that two linearly independent solutions of the wave equation will mix when they are transported around a non-trivial loop in the complex radial plane.  The failure of the field to be globally holomorphic is encoded in monodromy matrices associated with the singular points at the horizons and infinity.  At the horizons, these matrices have a simple physical interpretation: their eigenvectors correspond to the solutions that are purely ingoing/outgoing.  At infinity, ingoing/outgoing boundary conditions instead correspond to eigenvectors of a \emph{formal} monodromy matrix (as we will discuss in section \ref{sec:sing}). The scattering amplitude is then encoded in the change-of-basis matrix between two sets of monodromy eigenvectors.  The advantage of this procedure is that monodromy data can, in many cases, be extracted from purely {\it local} features of the geometry, without the need to evolve wave functions from the horizon to asymptotic infinity.  The topology of the complex radial plane (with singular points removed) leads to a non-trivial {\it global} constraint, allowing us to extract certain features of scattering data without ever solving a differential equation.

Although elegant, this algebraic procedure does not completely fix the reflection and transmission coefficients.  In simple backgrounds, such as BTZ black holes, it allows us to compute the norm of the transmission and reflection coefficients---i.e., the greybody factor---but not their phases.  The computation of quasinormal modes in these simple cases, then, only requires a little more work.
For more complicated backgrounds, such as Schwarzschild or Kerr black holes, the singular point associated with asymptotic infinity is irregular, so it exhibits Stokes phenomena; this  allows global features of the black hole geometry to sneak into monodromy data at infinity.  Describing these data thus requires either perturbative or numerical computations, which we develop.  Beyond this complication, as will become clear, there is another subtlety that prevents us from precisely computing scattering data in this manner.  Nevertheless, our method provides new analytic insight.  For example, quasinormal frequencies are expressed as roots of a certain transcendental equation, allowing us to qualitatively understand their analyticity properties in the complex frequency plane.

In the following section, we describe the monodromy technique, focusing on the extraction of scattering data and a treatment of Stokes phenomena.  In section \ref{sec:exkerr}, we apply this to the Kerr black hole.  In section \ref{sec:connmatrix}, we consider the symmetries of the relevant wave equation which, when combined with monodromy techniques, allow us to develop new methods for the study of Kerr quasinormal modes.  Many of the technical results are relegated to a series of appendices. We also refer the reader to a companion paper \cite{Castro:2013kea}, which relates this technique to microscopic treatments of black hole quantum mechanics in terms of a conformal field theory.


\section{Overview of the Monodromy Technique}
\label{sec:review}

Our goal is to study the dependence of scattering coefficients and quasinormal modes on analytic (global) properties of linearized fluctuations around a black hole background in asymptotically flat spacetime (other asymptotics can be treated in the same vein, with appropriate modifications).  A typical scattering computation requires one to solve a wave equation with specified boundary conditions.  The essential point is that when those boundary conditions are specified at singular points of the wave equation, then they will be intimately connected to analytic properties of solutions when the independent variable is complexified (typically the radial coordinate).  In this section, we will explain how knowledge of the analytic properties of the solutions can be used to compute scattering coefficients, reducing the computation to a simple exercise in linear algebra.

This section is rather technical; an eager reader more interested in black hole physics is invited to skim this section and head directly to section \ref{sec:exkerr}. A more complete treatment of the relevant ODEs can be found in, e.g., \cite{coddington1972theory,fokas2006painleve}.


\subsection{Preliminaries: ODEs and Monodromies}
\label{sec:ODEs}

We will focus on second order ordinary differential equations of the form
\be\label{aa:1}
\partial_z\big(U(z)\p_z\psi(z)\big) - V(z)\,\psi(z) =0~.
\ee
Such equations can always be rewritten as two coupled first order ODEs: for instance, introducing a two-component column vector $\Psi$ satisfying the ODE
\be\label{aa:2}
\partial_z \left(\begin{array}{c}\Psi_1\\ \Psi_2\end{array}\right) = \left(\begin{array}{cc}0 & \frac{1}{U(z)} \\ V(z) & 0 \end{array}\right) \left(\begin{array}{c}\Psi_1\\ \Psi_2\end{array}\right) =: A(z) \Psi ~,
\ee
which reproduces \eqref{aa:1} when we identify $\Psi_1$ with $\psi$ since then $\Psi_2$ equals $U(z)\p_z\psi$.  The space of solutions to this ODE is two-dimensional, so we can choose a linearly independent basis of solutions $\Psi^{(1)}$ and $\Psi^{(2)}$ and collect them into a \emph{fundamental matrix}
\be\label{aa:3}
\Phi(z) := \left(\begin{array}{cc}  \Psi^{(1)} & \Psi^{(2)} \end{array}\right)~.
\ee
The linear independence of the two solutions is equivalent to the invertibility of $\Phi(z)$.      

The Wronskian of two solutions, $\psi_1(z)$ and $\psi_2(z)$, is
\be
\label{eqn:Wronskian}
W(\psi_1,\psi_2) := U(z) \Big( \psi_1(z) \, \p_z \psi_2(z) - \psi_2(z) \, \p_z \psi_1(z) \Big) \, .
\ee
Note that the determinant of $\Phi(z)$, which equals the Wronskian of $\Psi_1^{(1)}$ with $\Psi_1^{(2)}$, is a constant since ${\rm tr}(A)=0$.  This follows from the path-ordered representation, $\Phi(z) = {\cal P}\big\{ \exp(\int^z A)\big\} \Phi_0$, which shows that $\det(\Phi(z)) = \det(\Phi_0)$ is independent of $z$ since
\be
 \det\Big( {\cal P}\big\{ e^{\int^z A} \big\} \Big) =  \det\Big( e^{\int^z A} \Big) = e^{\int^z {\rm tr}(A)} = 1 ~ ,
\ee
where $\mathcal{P}$ denotes path ordering.

We now analytically continue to the complex $z$-plane and consider cases where $A(z)$ is meromorphic; all cases considered here will certainly satisfy this requirement.  In this case, one may think of $A(z)$ as a flat $SL(2;\bb{C})$ connection,\footnote{Given a meromorphic $A(z)$, not necessarily of the form in \eqref{aa:2}, one might expect it would be a $GL(2;\bb{C})$ connection, however one can always perform a meromorphic gauge transformation to force $\textrm{tr}(A(z)) = 0$, as satisfied by \eqref{aa:2}.} where allowed gauge transformations must themselves be meromorphic functions of $z$.  The gauge-invariant information of $A(z)$, namely its holonomies, is thus encoded in the analytic structure of $\Phi(z)$; in particular, if $\Phi(z)$ is meromorphic,  $\Phi(z)$ \emph{is} the gauge transformation that sets $A(z)$ to zero.  To more directly see this relationship, follow $\Phi(z)$ around a closed loop $\gamma$ in the complex $z$-plane, calling the result $\Phi_\gamma(z)$.  Since $A(z)$ is meromorphic, the differential operator $\p_z - A$ returns to itself, which implies that $\Phi_\gamma(z)$ must again be a fundamental matrix for the ODE, however it need not be equal to $\Phi(z)$: given one fundamental matrix, we can always multiply it from the right by a constant invertible matrix to obtain another (i.e., we can choose a different linearly independent basis of solutions).  By the definition of a fundamental matrix, then, $\Phi_\gamma(z)$ must equal $\Phi(z)M_\gamma$ for some invertible matrix of constants $M_\gamma$:
\be
\label{eqn:monodef}
\Phi_\gamma(z) = \mathcal{P}\big\{ e^{\oint_\gamma A}\big\} \Phi(z) =: \Phi(z) M_\gamma \, .
\ee
 By definition, if our loop $\gamma$ does not cross any branch cuts of $\Phi(z)$ then $M_\gamma = \mathbf{1}$, so $M_\gamma$ is a measure of the lack of meromorphicity of $\Phi(z)$ and is called a \emph{monodromy matrix}.  A rearrangement of \eqref{eqn:monodef}
\be\label{eq:ma}
M_\gamma = \Phi(z)^{-1} \mathcal{P}\big\{ e^{\oint_\gamma A}\big\} \Phi(z)
\ee
emphasizes the relationship between the holonomy of $A(z)$ around $\gamma$ and the monodromy matrix $M_\gamma$: they are in the same conjugacy class.  

In particular, if one can find a gauge in which $A$ has no poles enclosed by $\gamma$, then $M_\gamma = \mathbf{1}$.  In this way, poles of $A$ that cannot be removed by gauge transformations correspond to branch points of $\Phi(z)$ and are associated with non-trivial monodromy matrices.  Since the conjugacy class of the holonomy $\mathcal{P}\big\{\!\exp{(\oint_\gamma A)}\big\}$ is independent of variations of the loop $\gamma$ that do not cross other branch points of $\Phi$, the monodromy matrices associated with the branch points define an embedding of the first homotopy group $\Pi_1\big(\mathbb{P}^1 \backslash\{\textrm{branch pts}\}\big)$ into $SL(2;\mathbb{C})$, where we have compactified the $z$-plane to a $\bb{P}^1$ by adding the point at infinity (which frequently is itself a branch point, as determined unambiguously by the connection $A$).

This has an implication that is key to the rest of our study.  Let $z=z_i$, for $i=1,\ldots,n$, be the locations of all branch points of $\Phi(z)$, and let $M_i$ be the monodromy matrix associated with a loop that encloses only the branch point at $z_i$.  If we follow $\Phi(z)$ around a path enclosing all branch points, the other side of the loop encloses no branch points and so the monodromy around that loop must be trivial.  In other words,
\be\label{aa:7}
M_1 M_2 \cdots M_n = \mathbf{1}~.
\ee
This innocent relation is actually quite interesting.  The conjugacy class of each individual $M_i$ can often be computed quite simply from \emph{local} information of the differential equation (with an important caveat in section \ref{sec:sing}), while \eqref{aa:7} is a relation among these local data---it is a piece of \emph{global} information.  Computing scattering coefficients is an example of a problem where we require global information about our solutions (relating boundary conditions at different points in the $z$ plane), so the rest of our work will explore how we may exploit this relation.

As we alluded to above, if $\Phi$ has a branch point at $z_i$, then $A$ has a pole there.  The converse is not true because it may be possible to remove the pole in $A$ by a gauge transformation.  Fortunately, there exists a simple algorithm for choosing gauge transformations to reduce the order of the pole to some minimal integral value called the \emph{Poincar\'e rank}, $\textsc{r}_i\in\mathbb{N}_0$, so there is a gauge where $A(z)$ takes the form
\be
\label{eq:minimal}
 (z-z_i)^{-\textsc{r}_i-1} \, A_0 (z) \, ,
\ee
where $A_0(z)$ has a convergent Taylor series expansion in some neighborhood containing $z=z_i$.  A simple pole $\textsc{r}_i=0$ is called a \emph{regular singular point}, while a higher order pole $\textsc{r}_i>0$ is called an \emph{irregular singular point} of rank $\textsc{r}_i$.

The distinction between regular and irregular singular points might seem artificial, but their implications for $\Phi$ are starkly different: regular singular points correspond to algebraic or logarithmic branch points in $\Phi$, while irregular singular points correspond to essential singularities in $\Phi$ and exhibit Stokes phenomenon.  Understanding regular singular points will suffice to illustrate the essence of our approach to scattering computations, so we focus first on this case in sections \ref{sec:regsing} and \ref{sec:scattering}, discussing the added complications of irregular singular points in section \ref{sec:sing}.


\subsection{Monodromies and Regular Singular Points}
\label{sec:regsing}

Our goal now is to find monodromies and the local behavior of $\Phi$ around singular points.  To find the conjugacy class of $M_i$, first perform a gauge transformation to bring $A$ to the minimal form \eqref{eq:minimal}, where now $\textsc{r}_i=0$.  For loops enclosing a single pole, this renders the path ordering trivial in the limit that the loop approaches the pole, but there can be one more subtlety: if the eigenvalues of ${\rm Res}_{z_i} (A)$ differ by nonzero integers, one must perform additional gauge transformations to make the eigenvalues equal (for a systematic procedure, see, e.g., \cite{coddington1972theory}).  Once that is done, the conjugacy class of the monodromy matrix \eqref{eq:ma} is easily determined to be 
\be
M_i \, \cong \,  \exp{\oint_{\gamma_i} (z-z_i)^{-1} A_0(z)} = e^{2\pi i A_0(z_i)} \ ,
\ee
where `$\cong$' denotes equal up to conjugation by an element of $SL(2,\mathbb{C})$.  It is then a general theorem that around $z=z_i$ we can write
\be
\label{eq:regular-series}
\Phi(z) \ = \ \left({ \sum_{n=0}^\infty} (z-z_i)^n \Phi_n\right) \, (z-z_i)^{N_i} \, ,
\ee
where the $\Phi_n$ are constant matrices with $\Phi_0$ invertible, and the series has a nonzero radius of convergence around $z_i$.  Following the solution around $z_i$ via $(z-z_i)\rightarrow e^{2\pi i}(z-z_i)$, we see that the monodromy matrix is
\be\label{eq:exp-mon-mat}
M_i = e^{2\pi i N_i} \ .
\ee
The conjugacy class of $N_i$ is thus the same as that of $A_0(z_i)$, so this can be easily read off the ODE (still assuming that $A$ has been put into minimal form around $z_i$).


\subsection{Monodromies, Boundary Conditions, and Scattering}
\label{sec:scattering}

The relationship between monodromies and choices of boundary conditions at the regular singular points---which is certainly a natural place to set boundary conditions for black holes where the horizon is a regular singular point---follows readily from \eqref{eq:regular-series}.  For simplicity, suppose that $M_i$ has distinct eigenvalues $e^{\mp 2\pi \alpha_i}$ (so $N_i$ has eigenvalues $\pm i \alpha_i$, up to shifts by integers), then we are free to choose our fundamental matrix to diagonalize $M_i$, in which case
\be\label{aa:regexpan}
\Phi(z) = \Big( \Phi_0 + O(z-z_i) \Big) \left(\begin{array}{cc} (z-z_i)^{i\alpha_i} & 0 \\ 0 & (z-z_i)^{-i\alpha_i} \end{array} \right) \ .
\ee
Approaching $z_i$ from a direction where ${\rm Im}\big(i\alpha_i\ln(z-z_i)\big)\neq0$, we see explicitly that one column corresponds to ingoing waves and the other to outgoing waves.  Diagonalizing the monodromy matrix $M_i$ therefore corresponds to choosing a basis with definite boundary conditions at $z_i$.\footnote{When $M_i$ has a nontrivial Jordan block (hence, equal eigenvalues), it seems more difficult to identify diagonalization of $M_i$ with ingoing and outgoing boundary conditions since one solution always has a logarithmic branch cut.  However, depending on the particular problem being studied, other choices of boundary conditions may be perfectly sensible, for example demanding regularity of the solution at $z_i$.}

A scattering computation often involves finding the change of basis between solutions that are ingoing/outgoing at one singular point and solutions that are ingoing/outgoing at another singular point (or, sometimes, normalizable/non-normalizable).  To be more concrete, focus on two singular points, $z_1$ and $z_2$, and two fundamental matrices $\Phi_1$ and $\Phi_2$ that diagonalize $M_1$ and $M_2$, respectively.  Since there are only two linearly independent solutions to the ODE, $\Phi_2^{-1}\Phi_1$ must be a constant matrix and, in fact,
\be\label{eq:conn-def}
{\cal M}_{2\to 1}=\Phi_2^{-1} \Phi_1
\ee
is nothing more than a change of basis from left-eigenvectors of $M_2$ to left-eigenvectors of $M_1$.  So far we have said nothing of the normalization of the various bases of solutions, so we are free to rescale the columns of $\Phi_1$ and $\Phi_2$, meaning the change of basis ${\cal M}_{2\to 1}$ is thus far defined only up to multiplication by a diagonal matrix on either side
\be
\label{eq:diagonal}
{\cal M}_{2 \to 1} \ \sim \ \left(\begin{smallmatrix} d_1 &  \\  & d_2 \end{smallmatrix}\right) {\cal M}_{2 \to 1}  \left(\begin{smallmatrix} d_3 &  \\  & d_4 \end{smallmatrix}\right) \ .
\ee
If, for example, we have a signature $(1,1)$ inner product for our solutions---the Klein--Gordon inner product in the case we will study---then we can choose the columns of $\Phi_1$ and of $\Phi_2$ to be normalized, lifting (most of) the rescaling ambiguity by forcing ${\cal M}_{2\to 1} \in SU(1,1)$:
\be\label{aa:99}
{\cal M}_{2 \to 1}= \left(\begin{array}{cc} \frac{1}{{\cal T}}  &  \frac{{\cal R}}{{\cal T}} \\ \frac{{\cal R}^*}{{\cal T}^*} & \frac{1}{{\cal T}^*} \end{array}\right)~,\qquad |{\cal R}|^2+|{\cal T}|^2=1~,
\ee
where ${\cal R}$ and ${\cal T}$ are the reflection and transmission coefficients, respectively.  The only remaining ambiguity will be in changing the normalizations by phases, but these will fortunately cancel when computing $|{\cal R}|$ or $|{\cal T}|$.

We have now seen how monodromy matrices relate to boundary conditions at regular singular points and we have seen how to compute the conjugacy class of the corresponding monodromy matrices.  If we knew the actual form of the monodromy matrices rather than just their conjugacy classes, computing the scattering matrix would reduce to the simple linear algebra problem of finding the change of basis between the left-eigenvectors of $M_2$ and the left-eigenvectors of $M_1$, then using the relation \eqref{eq:diagonal} to fix ${\cal M}_{2\to 1}$ to be an $SU(1,1)$ matrix.  In fact, there are many physically interesting situations where knowing the conjugacy classes and the relation \eqref{aa:7} allows one to reconstruct the matrices in a common basis.  For example, when there are two singular points then the monodromy matrices are inverses of each other and the scattering matrix is the identity. However the nature of the singular point is important, so \emph{don't be misled here}: for example, the Coulomb potential has two singular points, yet scattering is nontrivial because infinity is an irregular singular point and the choice of plane-wave boundary conditions at the irregular singular point does not diagonalize the associated monodromy matrix, as discussed in section \ref{sec:sing}.

A more interesting situation, relevant to our discussions of black holes, is when there are three singular points.  In this case, knowledge of the conjugacy class of $M_1$, $M_2$, and $M_3$, together with the global relation in \eqref{aa:7}, is enough to reconstruct the matrices themselves in a common basis.  Explicitly, given
\be
\det (M_i) =1\, , \qquad {\rm tr}( M_i) = 2\cosh(2\pi  \alpha_i)\, , \qquad  M_i \neq \mathbf{1} \, ,  \qquad  \textrm{for}~~~ i=1,2,3\, ,
\ee
and equation \eqref{aa:7},
\be
M_1 M_2 M_3 \ = \ \mathbf{1}~,
\ee
then a common basis is given by
\begin{gather}
 M_1=\left(\begin{array}{cc}0&-1\\ 1&2\cosh(2\pi  \alpha_1)\end{array}\right) ~,\qquad\qquad M_2= \left(\begin{array}{cc}2\cosh(2\pi  \alpha_2) &e^{2\pi  \alpha_3}\\ -e^{-2\pi  \alpha_3}&0\end{array}\right)~,  \nonumber \\  \label{eq:cbcb}
M_3=\left(\begin{array}{cc}e^{2\pi  \alpha_3}&0\\ 2\big(e^{-2\pi  \alpha_3}\cosh(2\pi  \alpha_1)-\cosh(2\pi  \alpha_2) \big)&e^{-2\pi  \alpha_3}\end{array}\right)~.
\end{gather}
We can readily write down a change of basis from left-eigenvectors of $M_2$, diagonalizing $M_2$ as ${\rm diag}\{e^{-2\pi\alpha_2},e^{2\pi\alpha_2}\}$, to left-eigenvectors of $M_1$, diagonalizing $M_1$ as ${\rm diag}\{e^{-2\pi\alpha_1},e^{2\pi\alpha_1}\}$, as
\be
\label{eq:scattering}
{\cal M}_{2\to 1} \ \sim \  \left(\begin{array}{cc} \sinh\pi(\alpha_2-\alpha_1+\alpha_3) & \sinh\pi(\alpha_2+\alpha_1+\alpha_3)  \\  \sinh\pi(\alpha_2+\alpha_1-\alpha_3)  &  \sinh\pi(\alpha_2-\alpha_1-\alpha_3)  \end{array}\right) \ .
\ee
Whether this can be made an $SU(1,1)$ matrix depends on details of the $\alpha_i$, however when it is possible---for instance, when the $\alpha_i$ are all real and ${\cal M}_{2\to 1}$ is invertible---we can read off the norm of the transmission coefficient without even computing the required diagonal transformation \eqref{eq:diagonal}:
\be
\label{eq:simple-transmission}
|{\cal T}|^2 = 1 - |{\cal R}|^2  =  \frac{\sinh(2\pi\alpha_1)\,\sinh(2\pi\alpha_2)}{\sinh\pi(\alpha_3+\alpha_1-\alpha_2)\, \sinh\pi(\alpha_3-\alpha_1+\alpha_2)} \ .
\ee

Of course, ODEs with three regular singular points have hypergeometric functions as solutions and are therefore well understood, and we have verified that these formulas are correct.  The challenge for applying the same methods to scattering off black holes becomes apparent in the next section where we discuss the consequences of an irregular singular point, which black hole backgrounds have at asymptotic infinity (a consequence of plane waves having essential singularities at infinity).  The basic idea for computing the scattering coefficients will be the same, but there will be additional steps and subtleties.


\subsection{Formal Monodromies are Fake: Irregular Singular Points}
\label{sec:sing}

Asymptotically flat black holes have irregular singular points at infinity.  This means that $\Phi(z)$ has an essential singularity at $z=\infty$, reflecting the fact that the solutions asymptote to plane waves.   

Although the black holes we will study in later sections have singularities of Poincar\'e rank 1, in this section we will consider general Poincar\'e rank $\textsc{r} \geq 1$.  For simplicity, let $A(z)$ have a rank $\textsc{r}\geq 1$ singularity at $z=\infty$, which means there is a gauge in which $A(z)$ can be written as
 \be
z^{\textsc{r}-1} A_0(z) \ ,  
\ee
where $A_0(z)$ has a convergent Taylor series expansion in $z^{-1}$ around $z=\infty$ and $A_0(\infty) \neq 0$ (furthermore, there is no gauge in which $A$ is less singular at $z=\infty$).  For simplicity, let $A_0(\infty)$ have maximal rank,\footnote{This is called the \emph{non-resonant} case, the \emph{resonant} case takes a bit more work, essentially requiring one to consider the ODE on some multi-sheeted cover of the $z$-plane.} and choose the gauge so that $A_0(\infty)$ is diagonal.  Then there exists a \emph{formal} fundamental matrix of the form
\be
\label{eqn:FFS}
\Phi_f(z) = P(z) e^{\Lambda(z)}~,
\ee
where $P(z)$ is a non-negative power series in $z^{-1}$ (generally \emph{not} convergent) and
\be\label{eq:essential-singularity}
\Lambda(z) = \sum_{a=1}^\textsc{r} \Lambda_a z^{a} - \Lambda_0 \log(z) \, ,
\ee
where the $\Lambda_a$ and $\Lambda_0$ are all constant diagonal matrices, determined by requiring $\Phi_f$ solve the ODE order by order in $z^{-1}$---in particular, $\Lambda_\textsc{r} = A_0(\infty)$.  It would seem that we can read off the monodromy from these solutions to be $e^{2\pi i \Lambda_0}$, as before (around infinity, $z\to e^{-2\pi i}z$ is the positive direction); however, \emph{this is not so} because the calculation is complicated by the fact that $P(z)$ is just an asymptotic series, not a convergent series.  For this reason, $e^{2\pi i\Lambda_0}$ is called the \emph{formal} monodromy, but we will shortly see its relation to the true monodromy.

For describing boundary conditions at $z=\infty$, though, we care about diagonalizing $\Lambda(z)$ (as opposed to the true monodromy) since this describes the dominant behavior of the solutions at $z=\infty$, telling us whether the solutions are ingoing or outgoing.  On the other hand, the quantity entering the product relation \eqref{aa:7} is the true monodromy, so it is crucial to understand the relationship between the two.  The distinction and relation between them is intertwined with the fact that solutions of ODEs around irregular singular points exhibit \emph{Stokes phenomenon}.  The defining feature of Stokes phenomenon is that it arises when one attempts to describe one function (e.g., a fundamental matrix of true solutions) in terms of a function with a \emph{different branching structure} (e.g., a formal fundamental matrix).  Again, this arises here because the formal solutions are generally not convergent series.

To give the general idea, let us be a bit schematic, clarifying the details in the following paragraphs.  Consider the following formal expression:
\be
\label{eq:formal-product}
\Phi_f(z)^{-1} \Phi_f(z) \ = \ e^{-\Lambda(z)} P(z)^{-1} P(z) e^{\Lambda(z)} \, .
\ee
The important observation to make is that we can only say that
\be
P(z)^{-1} P(z) \ \sim \ \mathbf{1}   \qquad (\textrm{as } z\to \infty) \, ,
\ee
in words, $P(z)^{-1}P(z)$ is asymptotic to $\mathbf{1}$ as $z\to \infty$.  This means it could differ from $\mathbf{1}$ by something with no series expansion around $z=\infty$, e.g., $e^{z}$, and this is precisely what happens.
If a product such as \eqref{eq:formal-product} were between two actual fundamental matrices rather than two formal ones, we would expect the result to be a constant matrix.  Since formal fundamental matrices are asymptotic to actual ones, the limit of \eqref{eq:formal-product} as $z$ tends to $\infty$ will be a constant matrix $S$, called a \emph{Stokes matrix}, with components
\be
S_{ij}  \ = \  \lim_{z\to \infty} e^{-\Lambda_{ii}(z)+\Lambda_{jj}(z)} \big( \delta_{ij} + O(z^{-1}) \big) \, ,
\ee
where the $O(z^{-1})$ terms are asymptotic to $0$ as $z\to \infty$.  Clearly, then, $S_{ii}=1$, but what about off-diagonal entries?  When we approach $z=\infty$ along a ray on which $\textrm{Re}\big((-\Lambda_{\textsc{r},ii}+\Lambda_{\textsc{r},jj})z^{\textsc{r}}\big) < 0$, the exponential forces $S_{ij}$ to vanish.  On the other hand, when $\textrm{Re}\big((-\Lambda_{\textsc{r},ii}+\Lambda_{\textsc{r},jj})z^{\textsc{r}}\big) > 0$, this may combine with the $O(z^{-1})$ terms to produce a finite result, implying that $S_{ij}$ has an upper or lower triangular structure.  This is most evident very near $\textrm{Im}\big((-\Lambda_{\textsc{r},ii}+\Lambda_{\textsc{r},jj})z^{\textsc{r}}\big) = 0$, where $\textrm{Re}\big((-\Lambda_{\textsc{r},ii}+\Lambda_{\textsc{r},jj})z^{\textsc{r}}\big)$ is maximal or minimal, as one can observe by studying the classic example of the Airy function.  This is called a \emph{Stokes ray} and represents the transition between two asymptotic expansions.\footnote{Around this ray, one of the asymptotic solutions is maximally larger than (dominant) the other (subdominant), and the errors obtained by truncating the expansion of the dominant one become of the same order as the subdominant solution.  This is classically phrased in terms of the coefficient of the subdominant solution changing discontinuously across the Stokes ray, but it has been shown that, in reality, the transition can be represented in a sharp but continuous fashion \cite{Berry-Smoothing}.}

Now we can be more precise.  Divide the neighborhood of the irregular singular point at $z=\infty$ into wedges,
\be
\Omega_{k} := \big\{ \ \tfrac{k\pi-\delta}{\textsc{r}} \ < \ \textrm{arg}(z) + \tfrac{1}{\textsc{r}}\textrm{Arg}(\Lambda_{\textsc{r},22} - \Lambda_{\textsc{r},11}) \ < \  \tfrac{(k+1)\pi+\delta}{\textsc{r}} \ \big\} \, ,  \qquad k\in\mathbb{Z}\, ,
\ee
where $0 < \delta \ll \pi$.  The preceding paragraph illustrates that we should only think of a given asymptotic expansion (e.g., one for which $P(\infty) = \mathbf{1}$) as being asymptotic to a given true solution within a wedge, $\Omega_k$.  Once we cross the next Stokes ray, the same true solution will have a different formal fundamental matrix (e.g., its asymptotic expansion in the next wedge will begin with $P(\infty)\neq \mathbf{1}$).  

Let $\Phi(z)$ be a true solution such that on $\Omega_0$ 
\be\label{eqn:def-pw-irr}
\Phi(z)\big|_{\Omega_0} \ \sim \ \Phi_f(z)\big|_{\Omega_0}  \qquad  (\textrm{as } z\to \infty\,, ~ z\in\Omega_0) \ .
\ee
To circle $z=\infty$ in the positive direction, we follow $\Phi(z)$ from $\Omega_0 \to \Omega_{-1} \to \ldots \to \Omega_{-2\textsc{r}}$.  Notice that along $\Omega_0 \cap \Omega_{-1}$, we have that ${\rm Re}\big(z^\textsc{r}(-\Lambda_{\textsc{r},11}+\Lambda_{\textsc{r},22})\big) > 0$, so the first Stokes matrix is upper triangular.  
This means that the subdominant solution, the first column of $\Phi$, may be added to the dominant solution, the second column of $\Phi$.  
Thus,
\be
\Phi(z) \big|_{\Omega_{-1}} \ \sim \ \Phi_f(z)\big|_{\Omega_{-1}} \left(\begin{smallmatrix} 1 & C_{0} \\ 0 & 1\end{smallmatrix}\right) =: \Phi_f(z)\big|_{\Omega_{-1}} S_0  \qquad (\textrm{as } z\to \infty,~ z\in\Omega_{-1}) \ .
\ee
with $C_0$ constant.

At the next overlap, $\Omega_{-1} \cap \Omega_{-2}$, the roles of dominant and subdominant solutions reverse, so the relevant Stokes matrix will be lower triangular:
\be
\Phi(z)\big|_{\Omega_{-2}} \ \sim \ \Phi_f(z)\big|_{\Omega_{-2}} \left(\begin{smallmatrix} 1 & 0 \\ C_{-1} & 1 \end{smallmatrix}\right) S_0 =: \Phi_f(z)\big|_{\Omega_{-2}} S_{-1} S_0  \qquad  (\textrm{as } z\to \infty,~ z\in\Omega_{-2}) \, ,
\ee
and the Stokes matrices $S_k$ continue alternating between upper and lower triangular.  Finally, $\Omega_{-2\textsc{r}} \cong \Omega_0$, so
\be
\Phi(z) \big|_{\Omega_{-2\textsc{r}}} \ \sim \ \Phi_f(z) \big|_{\Omega_{-2\textsc{r}}} S_{-2\textsc{r}+1} \cdots S_0   \ = \  \Phi_f(z) \big|_{\Omega_0} e^{2\pi i \Lambda_0} S_{-2\textsc{r}+1} \cdots S_0   \qquad ~ (\textrm{as } z\to \infty,~ z\in\Omega_{-2\textsc{r}})
\ee
where $e^{2\pi i \Lambda_0}$ is the formal monodromy.  This provides us with the identification of the true monodromy as
\be
M_\infty = e^{2\pi i \Lambda_0} S_{-2\textsc{r}+1} \cdots S_0 \ .
\ee
The hard work comes in determining the $C_k$, commonly called \emph{Stokes multipliers}. Their values are not solely determined by the local data of the singularity, they depend on all terms in the connection.  Some useful references with a more complete discussion than presented here are \cite{coddington1972theory,fokas2006painleve,le2006rank}. In appendix \ref{app:stokes}, we review a simple numerical method to compute $C_k$.  


\section{Example: Kerr Black Hole}
\label{sec:exkerr}

The details in the following section will focus on the study of a scalar field in the Kerr black hole background, but the methods are readily extendable to a broad class of physically relevant situations. We will first revisit the wave equation of the probe with particular emphasis on exploiting the machinery presented in section \ref{sec:review}. 


\subsection{Wave Equation}

Kerr is the ideal playground.  The solution contains the complexity needed to illustrate the method in the most simple manner.  To setup our notation and conventions, we start by reviewing aspects of the geometry. In Boyer--Lindquist coordinates, we have
\be\label{bb:geom}
ds^2={\Sigma\over {\Delta}}dr^2- {{\Delta}\over \Sigma}\left(dt -{a}\sin^2\theta\, d\phi\right)^2+ \Sigma d\theta^2+{ \sin^2\theta \over \Sigma} \left((r^2+{a}^2)\,d\phi-{{a}}\,dt\right)^2~,
\ee
where  
\be
\Delta=r^2 +{a}^2-2M r~,\qquad \Sigma=r^2+{a}^2\cos^2\theta~.
\ee
This describes a generic 4D black hole with mass $M$ and angular momentum $J=Ma$. The inner and outer horizons are located at
\be
{r}_\pm = M \pm \sqrt{M^2-{a}^2}~.
\ee

The Klein--Gordon equation for a massless scalar  is 
\be\label{bb:KG}
{1\over \sqrt{-g} } \partial_\mu\left(\sqrt{-g}g^{\mu\nu}\partial_\nu\psi\right)=0~.
\ee
Generalizations to massive probes and tensor fields is straightforward. Expanding in eigenmodes
\be\label{bb:eigen}
\psi(t,r,\theta, \phi) = e^{-i\omega t+im\phi}R(r)S(\theta)~,
\ee
and using \eqref{bb:geom}, makes equation \eqref{bb:KG}  separable. The spheroidal function $S(\theta)$ satisfies
\be\label{bb:angeqn}
\left[ {1\over \sin\theta}\partial_\theta
\left(\sin\theta\partial_\theta\right)-{m^2\over\sin^2\theta}+a^2\omega^2\cos^2\theta \right] S(\theta)=-K_\ell  S(\theta)~,
\ee
for eigenvalue $K_\ell$, and the radial equation for $R(r)$ is then
\bea\label{bb:10}
\left[
\partial_r \Delta \partial_r
+{({2Mr_+\omega}-{a}m )^2\over (r-r_+)(r_+-r_-)}   - {({2Mr_-\omega}-{a}m )^2\over (r-r_-)(r_+-r_-)}
 + (r^2+2M(r+2M) )\omega^2 \right] R(r)=K_\ell  R(r)~ .
\eea
Equations (\ref{bb:angeqn}) and (\ref{bb:10}) are of the confluent Heun type---each with two regular singular points and a rank-1 irregular singularity---coupled by a separation constant $K_\ell$.

Most of our discussion deals with the details of the radial equation.  Our interest in the angular equation \eqref{bb:angeqn} is solely to determine the eigenvalues $K_\ell$. This can be done systematically by treating the $\omega$-dependent  term as a (not necessarily small)  perturbation \cite{Press:1973zz,Abramowitz}. A basis of solutions is given by
\be\label{bb:angsoln}
S(\theta)= \sum_{\ell'} d_{\ell \ell'}^m(a\omega) Y_{\ell'}^m(\theta) ~,\qquad m=-\ell,\ldots,\ell\,, \quad~\ell=0,1,2,\ldots,
\ee
where $Y_{\ell}^m(\theta)$ are associated Legendre polynomials of the first kind. The coefficients $d_{\ell \ell'}^m(a\omega) $ are determined by a recursive relation obtained by replacing \eqref{bb:angsoln} in \eqref{bb:angeqn}.  A similar recursive method also determines $K_\ell$ to be
\be
K_\ell (a\omega)=\sum_{n=0}^\infty k_n\, (a\omega)^{2n}~,
\ee
where the first few terms are
\be
k_0 = \ell (\ell+1) ~,\quad   k_1= \frac{2m^2+1-2\ell(\ell+1)}{(2\ell-1)(2\ell+3)}~.
\ee
Alternatively, one can derive the coefficients $k_n$ using the monodromy technique explained in appendix \ref{sec:perturbation}. See appendix \ref{sec:angeqn} for the explicit computations.  There are also asymptotic expansions of $K_\ell$ for large $a\omega$; we refer the curious reader to \cite{Abramowitz} for further details. 


\subsection{Singular Points}\label{sec:singKerr}

We now focus our attention on global properties of equation \eqref{bb:10}. The branch points of solutions of the scalar wave equation for Kerr are located at $r=r_\pm,\infty$.  The singularities at the horizons, $r_\pm$, are regular singular points while the singularity at infinity is an irregular singular point of rank 1.  One can identify the nature of the singular points, for example, by following the steps in section \ref{sec:ODEs}.  As it turns out in this case, all singularities in the ODE \eqref{bb:10} are also branch points of the solutions.  Another equivalent way of determining the nature of the singularity at, say, $r=r_+$, would be to substitute a series expansion of the form
\be
\label{n1kerr}
R(r) = (r-r_+)^{\pm i\alpha_+} \big[ 1 + O\big(r-r_+\big)\big]
\ee
and solve for $\alpha_+$ and the coefficients of $O(r-r_+)$. In this case, then, $r_+$ corresponds to a regular singularity. A similar computation for the inner horizon at $r=r_-$ sets the value of the exponent $ \alpha_-$. We find that
\be\label{bb:residue}
\alpha_\pm := {{2Mr_\pm\omega}-{a}m \over r_+-r_-} ~.
\ee
So we now know the conjugacy classes of the monodromy matrices associated to the horizons:
\be\label{eq:matrices1}
M_{+}  \cong \left(\begin{array}{cc}e^{-2\pi  \alpha_+} &0\\ 0&e^{2\pi  \alpha_+} \end{array}\right) \, ,  \qquad ~~ M_{-}  \cong \left(\begin{array}{cc}e^{-2\pi  \alpha_-} &0\\ 0&e^{2\pi  \alpha_-} \end{array}\right)\, .
\ee

For irregular singular points of rank $\textsc{r}$, e.g., around $r=\infty$, one must also include exponential factors in the series expansion. For our case, where the singularity has rank 1 at $r=\infty$, the asymptotic expansion of the solutions to \eqref{bb:10} is of the form 
 \be
\label{n2kerr}
R(r) = e^{\mp i\omega r}r^{\mp i\lambda-1} \big[ 1 + O\big(r^{-1})\big]\,,
\ee
where $e^{\pm 2\pi\lambda}$ will correspond to the eigenvalues of the formal monodromy matrix, rather than to the eigenvalues of the true monodromy at infinity, $e^{\pm 2\pi\alpha_\irr}$, as we will now explain in detail.

\begin{figure}[t!]
\centering{
\includegraphics{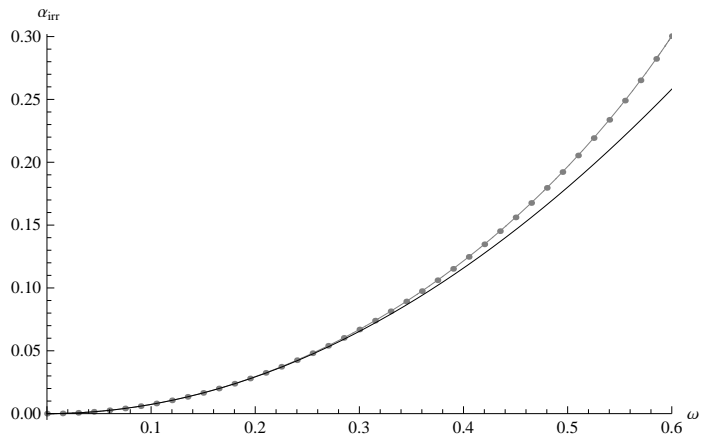}}
\caption{The figure depicts the monodromy around the irregular singular point, $\alpha_\irr$, as a function of the frequency $\omega$ for fixed values of the mass $M=0.7$, angular momenta $a=0.2$, and $\ell=m=2$. The {\it black} line outlines the analytic perturbative results for $\alpha_\irr$ while the {\it grey} line represents a fit to the numerical data given by the {\it grey dots}.}
\label{fig:alphainfty}
\end{figure}

At the singularity $r\to \infty$, there will be a nontrivial conjugacy class of the monodromy matrix of the form
\be\label{eq:matrices2}
M_{\infty}  \cong \left(\begin{array}{cc}e^{-2\pi  \alpha_\irr} &0\\ 0&e^{2\pi  \alpha_\irr} \end{array}\right) \, ,
\ee
where $\alpha_\irr$ is yet to be determined and will be more challenging to compute than for regular singular points. Since infinity is a (non-resonant) irregular singularity of rank 1, there are also two Stokes matrices that must be determined, which we denote by $S_k$.  Following the discussion in section \ref{sec:sing}, this implies that the monodromy at infinity can be expressed as 
\be\label{cc:5}
M_{\infty}= e^{2\pi i \Lambda_0} S_{-1} S_0 ~.
\ee
 The formal monodromy can be read off directly by replacing \eqref{n2kerr} in \eqref{bb:10}, and recalling that circling $r=\infty$ in a positive direction means taking $r\to e^{-2\pi i}r$,
\be
\label{eq:formal-monodromy}
e^{2\pi i\Lambda_0}  \cong \left(\begin{array}{cc}e^{-2\pi\lambda} &0\\ 0&e^{2\pi\lambda } \end{array}\right)~,\qquad \lambda=2 M\omega~.
\ee
It happens that $\lambda=\alpha_+-\alpha_-$ in the Kerr background, but we will only utilize this fact in our final expressions so that intermediate expressions are applicable to more general confluent Heun equations. In a basis where $e^{2\pi i \Lambda_0}$ is diagonal, we define
\be\label{bb:6}
 S_{-1}= \left(\begin{array}{cc}1 &0\\ C_{-1} &1\end{array}\right)  ~,\qquad  S_0= \left(\begin{array}{cc}1 & C_0\\ 0 &1\end{array}\right) ~,
\ee
where $C_{0,-1}$ are the Stokes multipliers.  Taking the trace  of \eqref{cc:5},  we find that
\be\label{cc:omega}
 \alpha_\irr= \frac{1}{2\pi} \, \cosh^{-1}\! \left[ \cosh(2\pi\lambda) + e^{2\pi\lambda }\, C_0 C_{-1} / 2 \right] \, .
\ee
Computing $\alpha_\irr$ is then equivalent to determining the product $C_0 C_{-1}$, which is a rather involved task.  Analytically, it is not obvious how to estimate $ \alpha_\irr$ for arbitrary values of the frequency. However, in appendix \ref{sec:perturbation} we develop a series expansion to compute $\alpha_\irr$ directly for low frequencies. Alternatively, Stokes multipliers can be computed numerically to high accuracy using the method summarized in appendix \ref{app:stokes}, developed by \cite{daalhuis1995calculation}. We developed the StokesNotebook \cite{StokesNotebook} (for Mathematica) that implements the aforementioned method to compute the Stokes multipliers and, hence, (\ref{cc:omega}) numerically. A comparison of the analytical and numerical results to order $O(\omega^4)$ in $K_{\ell}$ and $\alpha_\irr$ are shown in Fig. \ref{fig:alphainfty}.  Naturally, the perturbative approximation breaks down for larger values of $\omega$.

\subsection{Scattering Coefficients}
\label{sec:greybody}

For the purposes of this subsection, until we discuss quasinormal modes, the parameters in the radial wave equation \eqref{bb:10} are restricted to their typical physical ranges, i.e., $0\leq r_- < r_+$, $\omega \in \bb{R}$, $\ell=0,1,\ldots$, $m=-\ell,-\ell+1,\ldots,\ell$, and hence $K_\ell\in \bb{R}$.   
 When we discuss quasinormal modes later, the frequency $\omega$ will be extended to complex values.

Having identified the singular points, we are now ready to discuss reflection and transmission coefficients. As will be explicit below, global data (including Stokes data) is not sufficient to determine these physical observables. The derivations here are intended to quantify which data  are fixed by the monodromy information, and which are not. We will return to this computation in  section \ref{sec:connmatrix}.

We have three singular points---inner and outer horizons, and infinity---hence from equation \eqref{aa:7} we know that
\be\label{dd:0}
M_-M_+M_\infty=\mathbf{1}~.
\ee 
As we showed in section \ref{sec:scattering}, finding the relation between two bases of solutions defined by their behaviors at regular singular points is largely algebraic. Still, we should specify clearly the choice of boundary conditions and its relation to the discussion in section \ref{sec:scattering}. Therefore, 
our first task is to define ingoing and outgoing modes at the outer horizon and infinity, keeping the role of the monodromy as explicit as possible throughout the process.

As in any other scattering problem, the phase of the solution should have a definite sign as the boundary is approached to correspond to waves falling into or out of the horizon.  The ingoing/outgoing solutions at horizons are often written in so-called tortoise coordinates, $r_*$, where they take the form of plane waves:
 \be\label{pwbasis}
 \psi \sim e^{i\omega r_*}(\ldots)+  e^{-i\omega r_*}(\ldots)~,
 \ee
(note that even though the solutions have essential singularities at the horizon, $r_*=-\infty$, we need not worry about Stokes phenomenon since the ODE, expressed in tortoise coordinates, will not be meromorphic in $r_*$ and the machinery we have discussed is not directly applicable in tortoise coordinates).  For the outer horizon, adopting the left-eigenvectors of $M_+$ as a basis coincidences with ingoing/outgoing boundary conditions \eqref{pwbasis}, as the expansion in \eqref{n1kerr} indicates.  The fundamental matrix is therefore of the form
\bea
\Phi_+ &=& \bigg( \sum_{n=0}^{\infty} (r-r_+)^n \Phi_n^+ \bigg) \left(\begin{array}{cc} (r-r_+)^{i\alpha_+} & 0 \\ 0 & (r-r_+)^{-i\alpha_+} \end{array} \right) \ , \nonumber \\  \label{bb:baseouter}
&=& \left\{  \ \begin{array}{rcl}
\big(\begin{array}{cc}\Psi_{\rm out, +} & \Psi_{\rm in, +}\end{array}\big)\, , & \quad & \textrm{for}~~ \sgn\!\left(\omega\alpha_+\right) = + \,  \\
 \big(\begin{array}{cc}\Psi_{\rm in, +} & \Psi_{\rm out, +}\end{array}\big)\, , & \quad & \textrm{for}~~ \sgn\!\left(\omega\alpha_+\right) = - \, ,  \end{array}\right.  
\eea
where $\alpha_+$ is given by \eqref{bb:residue}, $\Phi_n^+$ are constant matrices, and $\Phi_0^+$ is invertible.  (When the parameters are instead complex, the sign of ${\rm Re}(\omega) {\rm Re}(\alpha_+)$ is the defining feature of ingoing versus outgoing.)

At infinity, the plane wave basis of the form \eqref{pwbasis} diagonalizes the formal monodromy $e^{2\pi i \Lambda_0}$. Hence the basis at infinity, valid in an open wedge of the complex $r$ plane that emanates from $r=\infty$ and contains the ray $r>0$, is of the form 
\bea\label{bb:pw}
\Phi_{\rm pw} &=&  \big(\begin{array}{cc}\Psi_{\rm in, \, pw} & \Psi_{\rm out, \, pw}\end{array}\big)\cr 
&=& \bigg( \sum_{n=0}^{\infty} r^{-n} \Phi_n^{\rm pw} \bigg) \left(\begin{array}{cc} e^{-i\omega r}r^{-i\lambda-1} & 0 \\ 0 &e^{i\omega r}r^{i\lambda-1} \end{array} \right)~.
\eea
Here $\lambda = 2M\omega$ as in \eqref{eq:formal-monodromy}, $ \Phi_n^{\rm pw}$ are constant matrices, $\Phi_0^{\rm pw}$ is invertible, and $\sum_n r^{-n} \Phi_n^{\rm pw}$ generally does not converge.

Next, we can use the Wronskian \eqref{eqn:Wronskian} to normalize solutions by defining an inner product on solutions via
\be\label{eq:www}
\langle \psi_1 , \psi_2 \rangle := \frac{W(\psi_1^*,\psi_2)}{2i\omega}= \frac{(r-r_+)(r-r_-)}{2i\omega} \, \Big( \psi_1^*(r) \, \p_r \psi_2(r) - \psi_2(r) \, \p_r \psi_1^*(r) \Big) \ .
\ee
This is a constant because $\psi_1^*$ will solve the same ODE as $\psi_1$ when the differential operator is real, as we have assumed for this subsection.  We then define the norm of a solution to be $|| \psi ||^2 := \langle \psi, \psi \rangle$, which is proportional to the flux through a surface of constant $r$.  In terms of this inner product, and under our assumptions on ranges of parameters (e.g., $r_+-r_->0$), we find that
\be
\sgn\big( ||\psi||^2 \big) = \left\{ \begin{array}{rcl} + & & \textrm{for `Out' states,} \\ - & & \textrm{for `In' states.} \end{array}\right.
\ee
Translating this definition into the language of monodromies gives the following. When a monodromy $\alpha$ is real, the basis diagonalizing the monodromy matrix can be normalized so that the inner product defined by the Wronskian,\footnote{As noted around \eqref{eqn:Wronskian}, recall that the Wronskian equals the determinant of $\Phi$.} which determines the flux through a fixed $r$ surface via the Klein--Gordon inner product, is given by $\sigma^3$.\footnote{Throughout, we will use $\sigma^i$ to denote the Pauli matrices.}  This is the case in the Kerr background for $\alpha_\pm$ at real frequencies, and it applies as well to the formal monodromy $\lambda$. 

The computation of greybody  factors (i.e., scattering coefficients) boils down to finding the connection matrix that transforms the two bases of solutions into one another. For our choice of boundary conditions at the horizon \eqref{bb:baseouter} and infinity \eqref{bb:pw}, the connection matrix is defined as
\be
{\cal M}_{{\rm pw}\to+}=\Phi_{\rm pw}^{-1}\Phi_+\ .
\ee
Since our basis can always be made orthonormal for $\alpha_+$ and  $\lambda$ real, the connection matrix can be brought to the form
\be\label{eq:mpw+}
{\cal M}_{{\rm pw} \to +}= \left(\begin{array}{cc} \frac{1}{{\cal T}}  &  \frac{{\cal R}}{{\cal T}} \\ \frac{{\cal R}^*}{{\cal T}^*} & \frac{1}{{\cal T}^*} \end{array}\right)~,\qquad |{\cal R}|^2+|{\cal T}|^2=1~.
\ee
Hence,  ${\cal M}_{{\rm pw}\to +}$ is an $SU(1,1)$ matrix, where  $|{\cal T}|$ and $|{\cal R}|$ are the transmission and reflection coefficient, respectively. 
To compute \eqref{eq:mpw+}, we will take advantage of the product relation \eqref{dd:0}. This product involves the true monodromy $M_\infty$ and not the formal monodromy, so we will split the computation into two parts: first we will connect $\Phi_+$ to the fundamental matrix $\Phi_\infty$ that diagonalizes $M_\infty$ via  \eqref{dd:0}, and then we will relate $\Phi_\infty$ to $\Phi_{\rm pw}$ using \eqref{cc:5}.

Note that the fundamental matrix that diagonalizes $M_\infty$ can be described in Floquet form as
\bea\label{bb:basetrue}
\Phi_\infty &=&  \big(\begin{array}{cc}\Psi_{\rm 1, \infty} & \Psi_{\rm 2, \infty}\end{array}\big)\cr 
&=& \bigg( \sum_{n=-\infty}^{\infty} r^n \Phi_n^\infty \bigg) \left(\begin{array}{cc} r^{-i\alpha_\irr} & 0 \\ 0 & r^{i\alpha_\irr} \end{array} \right) \ .
\eea    
In comparison with \eqref{bb:baseouter}, the expansion here is given by a Laurent series since $r=\infty$ is an irregular singularity.  We emphasize that this implies that $\alpha_\irr$ cannot be read off directly from the ODE.  

To convert from the basis of left-eigenvectors of $M_+$ to the plane-wave basis of left-eigenvectors of $e^{2\pi i \Lambda_0}$, we go through the intermediate basis diagonalizing $M_\infty$.  Following section \ref{sec:scattering}, the change of basis from $M_+$ to $M_\infty$ is given by
\be\label{bb:conp}
{\cal M}_{\infty\to +} \ \sim \  \left(\begin{array}{cc} \sinh\pi(\alpha_\irr-\alpha_+ +\alpha_-) & \sinh\pi(\alpha_\irr+\alpha_+ +\alpha_-)  \\  \sinh\pi(\alpha_\irr+\alpha_+ -\alpha_-)  &  \sinh\pi(\alpha_\irr-\alpha_+ -\alpha_-)  \end{array}\right) \, ,
\ee
where `$\sim$' denotes equal up to the equivalence relation \eqref{eq:diagonal}.  This relates the two bases via
  \be
 \Phi_+ =\Phi_\infty{\cal M}_{\infty\to+} \, .
  \ee
Through \eqref{cc:5}, the Stokes matrices tell us how to relate the basis of solutions diagonalizing the monodromy matrix $M_\infty$ to the plane wave basis that diagonalizes the formal monodromy $e^{2\pi i \Lambda_0}$.  The Stokes matrices are given in \eqref{bb:6} and the formal monodromy in \eqref{cc:5}.  From there, we can compute a change of basis between the two to be
\be
 \Phi_{ \infty} =\Phi_{\rm pw}{\cal M}_{{\rm pw} \to\infty} \, ,
  \ee
with
\be
\label{eq:pwtoinfty}
{\cal M}_{{\rm pw}\to \infty} \sim \left(\begin{array}{cc} e^{\pi  \alpha_\irr}&e^{-\pi  \alpha_\irr} \\
\sinh\pi({\lambda-\alpha_\irr}) & \sinh\pi({\lambda+\alpha_\irr}) 
 \end{array}\right)  \, ,
\ee
up to equivalence \eqref{eq:diagonal}.

From here we have
\be
{\cal M}_{{\rm pw}\to +}={\cal M}_{{\rm pw}\to \infty}{\cal M}_{\infty \to +}~,
\ee
with the caveat that we have to determine the normalization of the solutions to make the product of two connection matrices meaningful. In particular, we have to determine the normalization of the columns in $\Phi_\infty$ following \eqref{eq:www}. To start, for real frequencies, $e^{2\pi \alpha_\irr}$ can be either real or a phase.\footnote{When the ODE is real, if $\psi$ is a solution then the complex conjugate $\psi^*$ must be a solution as well.  Furthermore, if  $\psi$ has a definite monodromy $e^{2\pi\alpha_\irr}$ then $\psi^*$ has monodromy $e^{2\pi\alpha_\irr^*}$.  Since the two eigenvalues of $M_\infty$ are $e^{\pm2\pi\alpha_\irr}$, it must be the case that $e^{2\pi\alpha_\irr}$ is real or a phase.}
When $e^{2\pi \alpha_\irr}$ is real, then the freedom in the normalizations of solutions can be partially fixed by making ${\cal M}_{{\rm pw}\to\infty}$ and ${\cal M}_{\infty\to +}$ into $SU(1,1)$ matrices.  When it is a phase, the normalizations of the columns of $\Phi_\infty$ can be chosen so that the inner product will be given by $\sigma^2$.  In this case, we can require ${\cal M}_{{\rm pw}\to\infty}$ and ${\cal M}_{\infty\to +}$ to be determinant $1$ matrices that convert a $\sigma^3$ norm into a $\sigma^2$ norm, or vice versa.  In fact, requiring the matrices to be $SU(1,1)$ for real $e^{2\pi\alpha_\irr}$ will automatically result in the latter property when it is a phase.

Normalizing the columns of $\Phi_\infty$ in this way allows us to write
\be\label{cd:3}
{\cal M}_{{\rm pw}\to +} =     \, \smat d_1 & 0 \\ 0 & d_1^{-1} \esmat
\mat \frac{1}{{\cal T}_1} & \frac{{\cal R}_1}{{\cal T}_1} \\ \frac{{\cal R}_1}{{\cal T}_1} & \frac{1}{{\cal T}_1} \emat  
\smat d_2 & 0 \\ 0 & d_2^{-1} \esmat 
\mat \frac{1}{{\cal T}_2} & \frac{{\cal R}_2}{{\cal T}_2} \\ \frac{{\cal R}_2}{{\cal T}_2} & \frac{1}{{\cal T}_2} \emat  
\smat d_3 & 0 \\ 0 & d_3^{-1} \esmat \, ,
\ee
where we have defined
\be
{\cal R}_1 =\sqrt{e^{-2\pi\alpha_\irr}\frac{\sinh\pi(\lambda-\alpha_\irr)}{\sinh\pi(\lambda+\alpha_\irr)}} \ , \nonumber
\ee
\be\label{R2}
{\cal R}_2 = \sqrt{\frac{\sinh\pi(\alpha_++\alpha_-+\alpha_\irr)\,\sinh\pi(\alpha_+-\alpha_-+\alpha_\irr)}{\sinh\pi(\alpha_++\alpha_--\alpha_\irr)\, \sinh\pi(\alpha_+-\alpha_--\alpha_\irr)}} \ ,
\ee 
\be
{\cal T}_i^2 = 1-{\cal R}_i^2 \qquad {\rm for} \quad i=1,2. \nonumber
\ee
From the reality of the ODE---i.e., when the parameters such as $\omega$ are real, as we've assumed for this section---we deduce that the parameters $d_1$ and $d_3$ are phases while $d_2$ is a phase (respectively, real) when $e^{2\pi\alpha_\irr}$ is real (respectively, a phase).

With this parameterization, we can now read off the entries of ${\cal M}_{{\rm pw}\to+}$ as reflection and transmission coefficients.  The norm of the transmission coefficient is independent of $d_1$ and $d_3$ in \eqref{cd:3}.  However it does depend on the unknown function $d_2$, adding another layer of complexity on top of the computation of the true monodromy $\alpha_\irr$.  Explicitly, we have
\be
\label{eq:transmission}
|{\cal T}|^2 = 1-|{\cal R}|^2 = \frac{{\cal T}_1^2\,{\cal T}_2^2}{\big(d_2^2+{\cal R}_1{\cal R}_2\big)\big(d_2^{\,-2}+{\cal R}_1 {\cal R}_2\big)} \ .
\ee
In summary, we cast the transmission coefficient as a function of ${\cal R}_{1,2}$, which are specified in terms of the monodromies, and of the normalization $d_2$.  With the techniques presented here we cannot determine $d_2$ (or for that matter $d_{1,3}$). In principle this requires explicitly solving the ODE, which we have avoided. In section \ref{sec:connmatrix}, we will discuss further how to constraint analytic properties of $d_2$, while still avoiding solving the ODE.

The advantage of this approach is that it highlights the dependence of the scattering coefficients on the global properties of the solutions.  For one, it shows how the inner horizon data enters: even though we're studying a boundary-value problem on $r\in(r_+,\infty)$, the monodromy data of the inner horizon enters due to the global constraint \eqref{aa:7}.  This approach also appears to separate out the contributions that are intrinsic to the black hole from those that arise from the propagation to asymptotic infinity, and there is a sense in which this is true \cite{Castro:2013kea}.  Although, as we have seen, the undetermined parameter $d_2$ plays a central role in understanding the transmission coefficient, and presumably mixes the intrinsic data of the black hole with that of asymptotic infinity.  In section \ref{sec:connmatrix}, we will further constrain the dependence of $d_2$ on $\omega$, finding qualitative agreement with the low-frequency approximations.

\subsection{Quasinormal Modes}

Quasinormal modes (QNMs) represent resonances in black hole scattering problems and therefore yield important information about the spectrum of radiation from a black hole.  They are defined as solutions that are purely ingoing at the horizon $r=r_+$ and purely outgoing at infinity.  For real frequencies, it is not possible to satisfy these boundary conditions, so we now relax that condition and allow $\omega\in\bb{C}$.  This also means that the undetermined constants in \eqref{cd:3}, $d_1$, $d_2$, and $d_3$, will become complex numbers, rather than just phases.

In terms of the definitions in section \ref{sec:greybody}, we have
\be\label{cc:in-out}
\Psi_{\rm in,+} = \Psi_{\rm out, \, pw} \ ,
\ee
which, along with the condition $\det{\cal M}_{{\rm pw}\to+} = 1$, implies that the connection matrix for a quasinormal mode takes the form
\be\label{cc:connec}
{\cal M}^\qnm_{{\rm pw}\to+} = \mat \frac{1}{\cal T} & \frac{\cal R}{\cal T} \\ \frac{\cal R'}{\cal T'} & \frac{1}{\cal T'} \emat
= \mat 0 & -1  \\ 1 & \frac{1}{\cal T'} \emat
  \, ,\qquad  {\cal T'} \neq 0 \,  ,
\ee
 when ${\rm Re}(\omega){\rm Re}(\alpha_+) < 0$ (the columns should be interchanged for the other sign).  Since the frequency is complex, ${\cal T'}$ and ${\cal R'}$ are generally not complex conjugates of ${\cal T}$ and ${\cal R}$.  The quasinormal mode condition \eqref{cc:in-out}, or \eqref{cc:connec}, is viewed as a constraint on the frequency $\omega$; we denote those frequencies that obey \eqref{cc:connec} as $\omega_\qnm$.  The expression in equation (\ref{eq:transmission}) is still valid, but with ${\cal T'}$ and ${\cal R'}$ replacing ${\cal T}^*$ and ${\cal R}^*$, so ${\cal T}{\cal T'}$ diverges at $\omega= \omega_\qnm$.

From (\ref{cd:3}), we find explicitly
\be\label{eq:kk}
{\cal M}_{{\rm pw}\to +} = \frac{1}{{\cal T}_1{\cal T}_2}\mat  d_1 d_3 \left(d_2+d_2^{\,-1}{\cal R}_1 {\cal R}_2\right) & \frac{d_1}{d_3} \left(d_2 {\cal R}_2 + d_2^{\,-1}{\cal R}_1\right)   \\   
\frac{d_3}{d_1} \left(d_2 {\cal R}_1 + d_2^{\,-1}{\cal R}_2\right)  &  \frac{1}{d_1 d_3}\left(d_2 {\cal R}_1 {\cal R}_2 + d_2^{\, -1}\right)    \emat \ .
\ee
We want to bring this connection matrix to the form \eqref{cc:connec}. The first constraint is that the product of the off-diagonal  entries is equal to $-1$. Since the product is independent of $d_1$ and $d_3$, this can be rearranged to yield the equation
\be
\label{eq:offdiag}
\frac{1}{{\cal R}_1{\cal R}_2}
(d_2^{\,2} + {\cal R}_1 {\cal R}_2 )(d_2^{\,-2} + {\cal R}_1 {\cal R}_2)= 0 \ .
\ee
Similarly, the product of the diagonal entries is independent of $d_1$ and $d_3$ and must vanish (this is just the inverse of \eqref{eq:transmission}).  This implies
\be\label{eq:diag}
\frac{1}{{\cal T}_1^2 {\cal T}_2^2}  (d_2^{\,2} + {\cal R}_1 {\cal R}_2 )(d_2^{\,-2} + {\cal R}_1 {\cal R}_2) = 0 \ .
\ee
Together with \eqref{eq:offdiag}, this condition implies that $({\cal T}_1 {\cal T}_2)^2/({\cal R}_1 {\cal R}_2)$ either vanishes less rapidly than the left-hand side of \eqref{eq:offdiag}, or does not vanish at all.

Employing our results for the computation of the transmission coefficients \eqref{R2} and the data of the $\omega_\qnm$ from \cite{Berti:2009kk}, we can easily see that ${\cal T}_1{\cal T}_2$ and ${\cal R}_1{\cal R}_2$ are finite and nonzero.  Typical values of these constants can be found using StokesNotebook \cite{StokesNotebook};  for Schwarzschild  and Kerr black holes, see Tables \ref{tab:myfirsttable} and \ref{tab:mysecondtable} in appendix \ref{app:numKS}.  Appealing to the finiteness of ${\cal R}_1$ and ${\cal R}_2$ allows us to rewrite the condition \eqref{eq:offdiag} more succinctly as
\be\label{eq:qnms}
d_2^{\,2} + {\cal R}_1 {\cal R}_2 = 0  \qquad \textrm{or}  \qquad  d_2^{\,-2} + {\cal R}_1 {\cal R}_2 = 0 \ ,
\ee
naturally corresponding to the poles of the transmission coefficient \eqref{R2}. This computation makes it transparent that the spectrum of the black hole depends not only on the intrinsic data of the black hole, e.g., ${\cal R}_{1}$ and ${\cal R}_2$ which are determined by the monodromies, but also on knowledge of the solution along $r\in(r_+,\infty)$, which is encoded in  $d_2$. 

That the frequency $\omega$ should satisfy at least one of the relations in \eqref{eq:qnms} is a necessary, but not sufficient, condition for that frequency to correspond to a quasinormal mode.  The full set of solutions to \eqref{eq:qnms} correspond to those values of $\omega$ satisfying \emph{either} in-out or out-in boundary conditions.  To understand the appropriate subset corresponding only to in-out boundary conditions, we need to know the behavior of $(d_1 d_3)$ since its vanishing or divergence at solutions to \eqref{eq:qnms} can change whether the upper-left or lower-right entry of the scattering matrix vanishes.\footnote{A similar phenomenon happens for massive scalar fields in a BTZ background, which is solved by hypergeometric functions and can therefore be studied explicitly.  The analog of \eqref{eq:qnms} would tell us that we have two branches of solutions indexed by integers $n_1,n_2\in\mathbb{Z}$, and the restriction to in-out boundary conditions would then restrict $n_1$ and $n_2$ to be natural numbers.  If we wrote the scattering matrix in that case as
\begin{equation*}
\mat d_1 & 0 \\ 0 & d_1^{\, -1} \emat
\mat \frac{1}{{\cal T}_2} & \frac{{\cal R}_2}{{\cal T}_2} \\ \frac{{\cal R}_2}{{\cal T}_2} & \frac{1}{{\cal T}_2} \emat
\mat d_3 & 0 \\ 0 & d_3^{\, -1} \emat \ ,
\end{equation*}
with ${\cal R}_2$ and ${\cal T}_2$ given by \eqref{R2}, then we would find that $(d_1 d_3)$ would have poles or zeros at the solutions to the analog of \eqref{eq:qnms}.}
Without knowing $d_2$, it is difficult to see this explicitly. 
In the example below for the interior resonances, the structure of the branches of solutions will be more clear.

\subsubsection{Resonances in the Interior}

A completely different boundary-value problem would be to impose boundary conditions at the inner and outer horizons.  This setup has not attracted much attention, nevertheless it could be interesting for the purpose of studying the interior of the black hole, and since it is a boundary problem that we can solve precisely in our setup, it is worth discussing.

Consider the connection matrix relating the left-eigenvectors of the monodromy matrices $M_+$ and $M_-$.  Following our previous discussion (see section \ref{sec:scattering}), we have
\be
 \Phi_- =\Phi_+{\cal M}_{+\to-} ~,
\ee
with
\be\label{zz:conn}
{\cal M}_{+\to -} \ =  \smat d_4 & 0 \\ 0 & d_4^{-1} \esmat \left(\begin{array}{cc} \sinh\pi(\alpha_+-\alpha_-+\alpha_\irr) & \sinh\pi(\alpha_++\alpha_-+\alpha_\irr)  \\  \sinh\pi(\alpha_++\alpha_--\alpha_\irr)  &  \sinh\pi(\alpha_+-\alpha_--\alpha_\irr)  \end{array}\right)  \smat d_5 & 0 \\ 0 & d_5^{-1} \esmat \, ~,
\ee
where $d_{4}$ and $d_5$ are the undetermined normalizations of each basis.

Consider imposing quasinormal-mode-like conditions on ${\cal M}_{+\to -}$, given by \eqref{cc:connec};  i.e., we have $\Psi_{\rm in,-} = \Psi_{\rm out, \, +}$.  Requiring the matrix to be of the form \eqref{cc:connec} gives
\bea\label{zz:q1}
\alpha_+-\alpha_-+\alpha_\irr = i n_1 ~,\quad n_1\in \Z~, \cr
\alpha_+-\alpha_--\alpha_\irr = i n_2 ~,\quad n_2\in \Z~.
\eea
We see, too, that the normalizations $d_4$ and $d_5$ contain information about the resonances in the connection matrix since we must tune them to match the form in \eqref{cc:connec}. 
Depending on whether we choose in-out or out-in boundary conditions, there are two different branches of modes. The  boundary conditions in-out and out-in are translated to regularity conditions of the solutions at each asymptotic region. Let's assume that ${\rm Re} (i\alpha_+)>0$ and ${\rm Re} (i\alpha_-)<0$,  then the in-out boundary conditions will restrict the integers $n_{1,2}$ in \eqref{zz:q1}  to be positive, and the out-in boundary conditions will correspond to $n_{1,2}$ being negative.  

These boundary conditions are analogous to those imposed on the angular equation \eqref{bb:angeqn}; see also appendix \ref{sec:angeqn}. In the literature, these modes are known as spheroidal wave functions \cite{ronveaux1995}.


\section{Symmetries of the Connection Matrix }
\label{sec:connmatrix}

Having analyzed in detail the global properties of the ODE, we now turn to analyzing certain symmetries of these equations. By exploiting these very simple symmetries of the ODE, we will determine certain universal properties of the connection matrix ${\cal M}_{i\to j}$. We will mostly be interested in finding ${\cal M}_{i\to j}$ between a regular singular point and an irregular singular point, with the immediate applicability being to deduce further properties of the entries in \eqref{cd:3}. As we will show, this rather simple analysis has consequences for the spectrum of the black holes. 

To illustrate the construction, we first write explicit relations for the general class of confluent Heun equations (CHE) and then apply these to the problem we are interested in, namely the Kerr wave equations.  The analysis can certainly be applied to other classes of ODEs, as in the case of the double confluent Heun equation (see, e.g., \cite{ronveaux1995}) which is relevant for the Teukolsky equations in extremal Kerr.


\subsection{Symmetries of the Confluent Heun Equation}
\label{sec:Heun-symmetries}

This section follows \cite{kazakov}, using a more convenient notation for our purposes and correcting some typos.

Consider the confluent Heun equation, with regular singular points at $z=0$ and $z=1$, and an irregular singular point at $z=\infty$.  The regular singular points have monodromy eigenvalues $e^{\pm 2\pi \alpha_0}$ and $e^{\pm 2\pi \alpha_1}$, respectively.  The \emph{formal} monodromy eigenvalues at infinity will be called $e^{\pm 2\pi \lambda}$, and the structure of the essential singularities at infinity will be given by $e^{\pm i\varpi z}$.  In terms of these parameters, the confluent Heun differential operator can be written as
\bea\label{eqn:xx1}
\widehat{\mathcal{L}}_{[\alpha_0,\alpha_1,\lambda,\varpi,\kappa]}(z) &\!\!\!=\!\!\!&z (z-1\big) \Bigg[\frac{\p^2}{\p z^2} +\left( \frac{1-2i\alpha_0}{z}+\frac{1-2i\alpha_1}{z-1}+2i\varpi \right)\frac{\p}{\p z}    \\
&&  -\bigg( \frac{(\alpha_0+\alpha_1)^2+i(\alpha_0+\alpha_1)+\kappa}{z(z-1)} + \frac{i\varpi(2i\alpha_0+i\lambda-1)}{z}+\frac{i\varpi(2i\alpha_1+i\lambda-1)}{z-1}\bigg)\Bigg] \ . \nonumber
\eea
The parameter $\kappa$ is defined such that when $\varpi=0$---i.e., when the singularity at infinity is regular---then $-e^{\pm \pi i \sqrt{1+4\kappa}}$ are the monodromy eigenvalues.  All parameters are generically complex.\footnote{In relation to the Kerr radial equation \eqref{bb:10}, we have that $r=(r_+-r_-)z + r_-$, $\alpha_0 = \alpha_-$, $\alpha_1=\alpha_+$, $\lambda = 2M\omega$, $\varpi=\omega(r_+-r_-)$, and $\kappa=K_\ell+(a^2-8M^2)\omega^2$.  Kerr is not the most general CHE since $\lambda = \alpha_1 - \alpha_0$ and since the parameters ranges are restricted.}  With the choice of scaling in the equation, the asymptotic behavior of solutions to $\widehat{\cal L}_{[\alpha_0,\alpha_1,\lambda,\varpi,\kappa]}(z) \cdot \psi(z) = 0$ are
\bea
\psi(z) & \sim & z^{2i\alpha_0}\, , ~~~\textrm{\underline{or}}~~~ 1  \qquad (z\rightarrow 0) \ ,  \nonumber \\
\psi(z) & \sim & (1-z)^{2i\alpha_1}\, , ~~~\textrm{\underline{or}}~~~ 1 \qquad (z\rightarrow 1)  \ , \\
\psi(z) & \sim & e^{-2 i\varpi z} z^{i\alpha_0 + i\alpha_1 -i\lambda - 1}\, , ~~~\textrm{\underline{or}}~~~z^{i\alpha_0+i\alpha_1+i\lambda - 1}  \qquad (z\rightarrow \infty) \ . \nonumber
\eea
We could rescale $\psi(z) = z^{i\alpha_0}(1-z)^{i\alpha_1} e^{-i\varpi z } \tilde{\psi}(z)$, in which case $\tilde{\psi}(z)$ would have more symmetric asymptotic expansions around the singular points, but we keep the current scaling to more simply connect with the notation in \cite{kazakov}.  When $\varpi=0$, the asymptotic expansion of solutions is convergent, and the leading behavior of an asymptotic expansion of the solutions around $z=\infty$ is
\be
z^{i\alpha_0 + i\alpha_1 \pm \frac{1}{2}\sqrt{1+4\kappa}-\frac{1}{2}} \ .
\ee
On the other hand, when $\varpi\neq 0$, the asymptotic expansion of solutions around $z=\infty$ is generically not convergent and $\kappa$ does not appear at leading order in the asymptotic expansion (not surprisingly, since the operations of expansion around $z=\infty$ and taking the limit $\varpi\rightarrow 0$ need not commute).

Let us first list the elementary symmetries of the confluent Heun equation, which relate the connection coefficients of two different confluent Heun equations to each other:
\bea\label{cm:aa}
\widehat{\mathcal{L}}_{[\alpha_0, \alpha_1, \lambda, \varpi, \kappa]}(z) \cdot \big( z^{2i\alpha_0} \tilde\psi(z) \big) = 0 \quad &\Longrightarrow& \quad 
\widehat{\mathcal{L}}_{[-\alpha_0, \alpha_1, \lambda, \varpi, \kappa]}(z) \cdot \tilde\psi(z)  = 0 \ , \\
\label{cm:ab}
\widehat{\mathcal{L}}_{[\alpha_0, \alpha_1, \lambda, \varpi, \kappa]}(z) \cdot \big( (1-z)^{2i\alpha_1} \tilde\psi(z) \big) = 0 \quad &\Longrightarrow& \quad \widehat{\mathcal{L}}_{[\alpha_0, -\alpha_1, \lambda, \varpi, \kappa]}(z) \cdot \tilde\psi(z)  = 0 \ ,  \\
\label{cm:ac}
\widehat{\mathcal{L}}_{[\alpha_0, \alpha_1, \lambda, \varpi, \kappa]}(z) \cdot \big( e^{-2 i\varpi z } \tilde\psi(z) \big) = 0  \quad &\Longrightarrow &\quad \widehat{\mathcal{L}}_{[\alpha_0, \alpha_1, -\lambda, -\varpi, \kappa]}(z) \cdot \tilde\psi(z)  = 0 \ .
\eea
Denote a fundamental matrix that diagonalizes the monodromy matrix at $z=z_i$ by $\Phi_{z_i}(z)$,\footnote{Note that the choice of $\Phi_{\rm pw}$ defined in \eqref{eqn:def-pw-irr} and section \ref{sec:greybody} does not diagonalize the monodromy matrix $M_\infty$, rather it diagonalizes the formal monodromy \eqref{cc:5}. Hence $\Phi_{\rm pw}$ diagonalizes $e^{2\pi i \Lambda_0}$, while $\Phi_{0},$ $\Phi_{1}$, and $\Phi_\infty$, diagonalize $M_{0}$, $M_{1}$, and $M_\infty$, respectively.} and the connection matrix \eqref{eq:conn-def} between any two pairs of such fundamental solutions by ${\cal M}_{z_i\to z_j}$. The elementary symmetries \eqref{cm:aa} and \eqref{cm:ab} imply that the connection matrix between two regular points must be of the form\footnote{One can also choose to normalize the fundamental matrices such that ${\cal M}_{i\to j}$ will have unit determinant.}
\be\label{cm:bb}
{\cal M}_{1\to 0}= \Phi_1^{-1}\Phi_0 =  \left(\begin{array}{cc} k(\alpha_0,\alpha_1,\lambda,\varpi,\kappa) & k(-\alpha_0,\alpha_1,\lambda,\varpi,\kappa)  \\ k(\alpha_0,-\alpha_1,\lambda,\varpi,\kappa)  & k(-\alpha_0,-\alpha_1,\lambda,\varpi,\kappa)  \end{array}\right)~.
\ee
 The same function $k$ determines all entries of the connection matrix. Invariance under \eqref{cm:ac} of $\Phi_{0}$ and $\Phi_{1}$ implies that
\be
k(\alpha_0,\alpha_1,\lambda,\varpi,\kappa) =k(\alpha_0,\alpha_1,-\lambda,-\varpi,\kappa) ~.
\ee
Next, consider ${\cal M}_{{\rm pw} \to 1}$, which relates the fundamental matrix diagonalizing $M_1$ to the formal fundamental matrix diagonalizing the formal monodromy around the irregular singular point $z=\infty$, \eqref{cc:5}. In a similar fashion, we can use \eqref{cm:ab} and \eqref{cm:ac} to constrain the entries of the matrix:
\be
{\cal M}_{{\rm pw}\to 1}= \Phi_{\rm pw}^{-1}\Phi_1 =  \left(\begin{array}{cc} q(\alpha_0,\alpha_1,\lambda,\varpi,\kappa) & q(\alpha_0,-\alpha_1,\lambda,\varpi,\kappa)  \\ 
q(\alpha_0,\alpha_1,-\lambda,-\varpi,\kappa)  &q(\alpha_0,-\alpha_1,-\lambda,-\varpi,\kappa)  \end{array}\right)~.
\ee
The transformation \eqref{cm:aa} further  implies that 
\be
q(\alpha_0,\alpha_1,\lambda,\varpi,\kappa) =q(-\alpha_0,\alpha_1,\lambda,\varpi,\kappa) ~.\ee

Another elementary symmetry arises from reflection $z=1-u$, which switches the two regular singular points. Call the generator of the reflection $\mathsf{r}$.  Let $\tilde\psi(u):=\psi(1-u)$, then
\be
\widehat{\mathcal{L}}_{[\alpha_0, \alpha_1, \lambda, \varpi, \kappa]}(1-u) \cdot \psi(1-u) = 0  \quad \Longrightarrow \quad \widehat{\mathcal{L}}_{[\alpha_1, \alpha_0, \lambda, -\varpi, \kappa]}(u) \cdot \tilde\psi(u)  = 0 \ .
\ee
The reflection symmetry $\mathsf{r}$ has a particularly interesting implication.  Consider how reflection acts on the infinite cover of the $z$-plane around the branch point at $z=\infty$ by writing $v=\frac{2}{2z-1} = \rho e^{i\theta}$, where $\theta\in\mathbb{R}$.  Then $z\rightarrow 1-z$ sends $v\rightarrow - v$, which means $\theta \rightarrow \theta + (2n+1)\pi$ for some $n\in\bb{Z}$.  This $\bb{Z}_2$ reflection, then, apparently lifts to an action of $\bb{Z}$ on the infinite-sheeted cover of the $z$-plane.  In that case, we should consider the ``reflection'' operator $\mathsf{r}$ to act as $\mathsf{r}:\, (z,\varpi)\rightarrow (1+e^{i\pi}z,e^{-i\pi}\varpi)$ on the cover of the $z$-plane.

Since $\Phi_0$ and $\Phi_1$ have convergent series expansions, we can act with $\mathsf{r}$ term-by-term on their series expansions and deduce that
\bea
\mathsf{r} \cdot \Phi_{1,[\alpha_0, \alpha_1, \lambda, \varpi, \kappa]}(z)  &=& \Phi_{0,[\alpha_1, \alpha_0, \lambda, -\varpi, \kappa]}(z) \, , \nonumber \\
\mathsf{r} \cdot \Phi_{0,[\alpha_0, \alpha_1, \lambda, \varpi, \kappa]}(z)  &=& \Phi_{1,[\alpha_1, \alpha_0, \lambda, -\varpi, \kappa]}(z) \, , \nonumber \\
\mathsf{r}^2 \cdot \Phi_{0,[\alpha_0, \alpha_1, \lambda, \varpi, \kappa]}(z)  &=& \Phi_{0,[\alpha_0, \alpha_1, \lambda, \varpi, \kappa]}(z) \, , \\
\mathsf{r}^2 \cdot \Phi_{1,[\alpha_0, \alpha_1, \lambda, \varpi, \kappa]}(z)  &=& \Phi_{1,[\alpha_0, \alpha_1, \lambda, \varpi, \kappa]}(z) \, . \nonumber
\eea
Thus, acting with $\mathsf{r}$  on \eqref{cm:bb} implies that the connection matrix must satisfy
\be
{\cal M}_{1\to 0}(\alpha_0,\alpha_1,\lambda,\varpi,\kappa) {\cal M}_{1\to 0}(\alpha_1,\alpha_0,\lambda,-\varpi,\kappa) =\mathbf{1}_{2\times 2}~.
\ee
On the other hand, the action on $\Phi_{\rm pw}$ cannot be naively deduced from the asymptotic expansion since it is not convergent and since the operation $\mathsf{r}^2$ takes us to a different sheet of the $z$-plane.  Instead, we see that
\bea
\mathsf{r}^2 \cdot \Phi_{{\rm pw},[\alpha_0, \alpha_1, \lambda, \varpi, \kappa]}(z)  &=& \Phi_{{\rm pw},[\alpha_0, \alpha_1, \lambda, e^{-2\pi i}\varpi, \kappa]}\big(e^{2\pi i}z\big)   \nonumber \\
&=& \Phi_{{\rm pw},[\alpha_0, \alpha_1, \lambda, e^{-2\pi i}\varpi, \kappa]}(z) \, M_{\infty,[\alpha_0, \alpha_1, \lambda, e^{-2\pi i}\varpi, \kappa]}^{-1} \\
&=& \Phi_{{\rm pw},[\alpha_0, \alpha_1, \lambda, \varpi, \kappa]}(z) \, M_{\infty,[\alpha_0, \alpha_1, \lambda, \varpi, \kappa]}^{-1} \, , \nonumber
\eea
where in the last line we have used the fact that $\Phi_{\rm pw}$ and $M_\infty$ are analytic in $\varpi$.\footnote{Since $\Phi_{\rm pw}$ is uniquely defined by its asymptotic expansion in a particular wedge of the $z$-plane, and since the coefficients of this asymptotic expansion will be analytic in $\varpi$, $\Phi_{\rm pw}$ is analytic in $\varpi$.  By the definition of the monodromy matrix, analyticity of $\Phi_{\rm pw}$ with respect to $\varpi$ implies analyticity of $M_\infty$.}

In that case, then, there are interesting implications for the connection matrix between a regular and irregular singular point.  Acting with $\mathsf{r}^2$ on
\be
\Phi_{1,[\alpha_0, \alpha_1, \lambda, \varpi, \kappa]}(z) = \Phi_{{\rm pw},[\alpha_0, \alpha_1, \lambda, \varpi, \kappa]}(z) \, \mathcal{M}_{{\rm pw}\rightarrow 1}\big(\alpha_0, \alpha_1, \lambda, \varpi, \kappa\big) \ ,
\ee
 implies that
\be\label{cm:cc}
\mathcal{M}_{{\rm pw}\rightarrow 1}\big(\alpha_0, \alpha_1, \lambda, e^{-2\pi i}\varpi, \kappa\big) = M_{\infty,[\alpha_0, \alpha_1, \lambda, \varpi, \kappa]} \mathcal{M}_{{\rm pw}\to 1}\big(\alpha_0, \alpha_1, \lambda, \varpi, \kappa\big) \,   ,
\ee
where $M_\infty$ is expressed in the plane-wave basis.  So we see that the connection matrix has a branch cut in the complex $\varpi$-plane around $\varpi=0$ with the same monodromy as $\Phi_{\rm pw}$ has around $z=\infty$.  Writing $M_{\infty}$ as
\be
M_\infty = e^{2\pi i N_\infty} \, ,
\ee
(see equation \eqref{eq:exp-mon-mat}), then a solution to \eqref{cm:cc} is
\be
\label{eq:connect-analyticity}
\mathcal{M}_{{\rm pw}\to 1} = \varpi^{-N_\infty} \mathcal{M}(\alpha_0, \alpha_1, \lambda, \varpi, \kappa)  \, ,
\ee
where $\mathcal{M}$ is a meromorphic function of $\varpi$. The connection matrix is not a single-valued function of $\varpi$, and the branching structure is dictated by the monodromy at infinity via $N_\infty$. 

The branch cut in the $\varpi$-plane that we point out here occurs in other contexts, as well.  For example, it is also known to occur in the double confluent Heun equation  \cite{ronveaux1995} (relevant to extremal Kerr),  and even in the case of confluent hypergeometric (relevant to the quantum mechanics of a charged particle in a Coulomb potential)---in the latter case, it will not manifest itself in the transmission coefficient ${\cal T}{\cal T}^*$ for reasons we point out in section \ref{sec:KR}.

By exploiting these very simple symmetries of the CHE, we were able to extract universal properties of the connection matrix. As we will discuss in the following sections, these properties have implications on the spectrum of the black hole. 
In addition,  there are also various interesting integral relations in \cite{kazakov} that can be exploited to further constrain the entries of ${\cal M}_{z_i\to z_j}$, though we will not use them here.


\subsection{Kerr Revisited}\label{sec:KR}

As an application of the above discussion, let's see how these results constrain our expressions for the greybody factors and quasinormal modes of the Kerr wave equation.  In section \ref{sec:greybody}, we constructed the connection matrix, which in \eqref{cd:3} we expressed as
\be\label{eq:44}
{\cal M}_{{\rm pw}\to +} = \smat d_1 & \\  & d_1^{\,-1} \esmat
 {\cal M}_{{\rm pw}\to\infty}    
\smat d_2 & \\  & d_2^{\,-1} \esmat
{\cal M}_{\infty\to +} 
\smat d_3 & \\  & d_3^{\,-1} \esmat \, ,
\ee
with  ${\cal M}_{\infty\to +} $ and ${\cal M}_{{\rm pw}\to\infty}$  given by \eqref{bb:conp} and \eqref{eq:pwtoinfty}, and where $d_i$ are undetermined parameters.

On the other hand, the analytic property of the scattering matrix that we just deduced in \eqref{eq:connect-analyticity} states that
\be
\label{eq:45}
\mathcal{M}_{{\rm pw}\rightarrow +} =  \varpi^{-N_\infty} \mathcal{M}(\alpha_-, \alpha_+, \lambda, \varpi, K_\ell) \, , 
\ee
where $\varpi = \omega (r_+ - r_-)$.  Since ${\cal M}_{{\rm pw}\to\infty}$ and ${\cal M}_{\infty\to +} $ are meromorphic functions of $\varpi$, the branch cut structure is necessarily encoded in the normalizations $d_i$ in \eqref{eq:44}.   To see this, first note that
\be
\smat d_1 & \\  & d_1^{\,-1} \esmat
 {\cal M}_{{\rm pw}\to\infty}
\ee
diagonalizes $M_\infty = e^{2\pi i N_\infty}$ from the plane-wave basis and, hence, diagonalizes $N_\infty$, too. So we can use \eqref{eq:44} and \eqref{eq:45} to write
\be
\mathcal{M}(\alpha_-, \alpha_+, \lambda, \varpi, K_\ell) = \smat d_1 & \\  & d_1^{\,-1} \esmat
 {\cal M}_{{\rm pw}\to\infty}    
\smat  \varpi^{i\alpha_\irr} & \\ & \varpi^{-i\alpha_\irr} \esmat
\smat d_2 & \\  & d_2^{\,-1} \esmat
{\cal M}_{\infty\to +} 
\smat d_3 & \\  & d_3^{\,-1} \esmat \, .
\ee
Since $\mathcal{M}(\alpha_-, \alpha_+, \lambda, \varpi, K_\ell)$ is meromorphic in $\varpi$, this implies that
\be\label{eq:d2d2}
d_2 = \varpi^{-i\alpha_\irr} d(\alpha_-,\alpha_+,\lambda,\varpi,K_\ell) \, ,
\ee
where $d(\alpha_-,\alpha_+,\lambda,\varpi,K_\ell)$ is meromorphic in $\varpi$.

The implications of the symmetry transformations on the transmission coefficient are straight forward. From \eqref{eq:transmission} we have 
\be
|{\cal T}|^2  = \frac{(1-{\cal R}_1^2) (1-{\cal R}_2^2)}{\big(d_2^2+{\cal R}_1{\cal R}_2\big)\big(d_2^{\,-2}+{\cal R}_1 {\cal R}_2\big)} \ ,
\ee
so we see that the scattering coefficients are clearly not analytic in $\omega$.  In particular, a low-frequency expansion of the non-analyticity of \eqref{eq:d2d2} gives
\be\label{eq:loww}
\varpi^{-2i\alpha_\irr} = \big(\omega(r_+-r_-)\big)^{2\ell}\Big[ 1  - (2M\omega)^2 \ln\!\big(\omega(r_+-r_-)\big) \Big( \tfrac{15\ell(\ell+1)-11}{(2\ell+3)(2\ell+1)(2\ell-1)}\Big) + O\big(\omega^3\ln\omega\big) \Big] \ ,
\ee
where we used the expansion of $\alpha_\irr$ in \eqref{eqn:monodromy}.  The leading power  of $\omega$ is in accord with results at low frequencies \cite{Staronbinski:1974,Teukolsky:1974yv,Page:1976ki}, with the caveat that we made a choice of branch in evaluating  $\alpha_\irr$ since $\cosh(2\pi\alpha_\irr)$ is really the well-defined quantity (so $\alpha_\irr$ is determined up to an overall sign and a shift by an imaginary integer).  This ambiguity can be fixed in several ways, the simplest being by requiring the decay rate to be finite in the limit that $M\omega\to 0$.\footnote{The relationship between the greybody factor $|{\cal T}|^2$ and the decay rate $\Gamma$ is simply $$\Gamma = {|{\cal T}|^2\over e^{4\pi \alpha_+}-1}~,\qquad \alpha_+={2M r_+ \omega-a m\over r_+ -r_- }~.$$}

A similar non-analyticity  in the transmission coefficients has been reported in various places, for example, in \cite{Hartle:1974,Sasaki:1989ca,Glampedakis:2001js,Mano:1996vt,Mano:1996gn}.  
In particular, the authors in \cite{Mano:1996vt,Mano:1996gn} constructed Floquet solutions at infinity (see \eqref{bb:basetrue}) and used them to match to the series solutions around the outer horizon.  The challenge in using Floquet solutions is twofold: one, it is difficult to compute the Laurent coefficients without knowing $\alpha_\irr$ exactly and without having enough boundary conditions for the coefficients, and two, the relationship to plane waves must be established to solve the desired boundary-value problem.  The advantage, though, is that the Laurent series is convergent.  The expressions the authors of \cite{Mano:1996gn} arrived at for the greybody factors are non-analytic functions of $\omega$ with the same branching structure we arrived at in \eqref{eq:d2d2}--\eqref{eq:loww} from our simple analysis of symmetries of the CHE.

Recall that the connection coefficient $d_2$ was particularly important in determining the location of quasinormal modes. With this new information, we  find that  \eqref{eq:qnms}
becomes
\be\label{eq:teqn}
\varpi^{-2i\alpha_\irr} d(\alpha_-,\alpha_+,\lambda,\varpi,K_\ell)^{2} + {\cal R}_1 {\cal R}_2 = 0   \qquad \textrm{or} \qquad  \varpi^{2i\alpha_\irr} d(\alpha_-,\alpha_+,\lambda,\varpi,K_\ell)^{-2} + {\cal R}_1 {\cal R}_2 = 0 \ .
\ee
Our simple analysis shows that the QNM spectrum of the Kerr black hole is governed by  a transcendental equation.  There are a few possibilities for solutions to this equation:
\begin{enumerate}
\item Both terms separately vanish in \eqref{eq:teqn}, in which case if $\omega$ is a QNM, then so is $e^{2\pi i}\omega$.  

\item When $\omega=\omega_\qnm$ is a QNM, $\alpha_\irr(\omega_\qnm)$ is rational and both terms cancel each other. Then, the QNMs live on a finite-sheeted cover of the $\omega$-plane.  

\item Quasinormal modes require the cancellation of the two terms against each other and each QNM lives on a single sheet of an infinite-sheeted cover of the $\omega$-plane ($\alpha_\irr(\omega_\qnm) \notin \mathbb{Q}$). 
\end{enumerate}
We can explicitly test whether the first two options are viable. It is clear from the data in Tables \ref{tab:myfirsttable} and \ref{tab:mysecondtable} that  $\alpha_\irr$ is generically complex, and ${\cal R}_{1}{\cal R}_2$ does not vanish for $\omega_\qnm$. 
The final option is the conclusion we are inevitably led to.

Further knowledge of how $d(\alpha_-,\alpha_+,\lambda,\varpi,K_\ell)$ in \eqref{eq:teqn} depends on its parameters is desirable.  This would lead to a better physical understanding of the properties of the spectrum, such as the spacing of the modes, it could lead to new symmetries in the spectrum, or it could provide a simple way to see that all QNMs decay in time (i.e., that ${\rm Im}(\omega_\qnm) < 0$).

The branch cut structure in the connection matrix can be attributed to the irregular singularity at infinity, making it difficult to disentangle which contributions of the QNM spectrum are intrinsic to the black hole.  In the low-frequency limit, we see that there is a  logarithmic contribution to the greybody that can be completely attributed to the presence of an irregular singularity \cite{Ching:1994bd,Ching:1995tj}.  Logarithmic branch cuts in the greybody factor have also been noticed, for instance, in certain zero-temperature correlation functions \cite{Sasaki:1989ca,Starinets:2002br}.  Since an extremal horizon generates an irregular singularity, this also would generate a logarithmic contribution to the greybody factor, fitting nicely with this zero-temperature result.  This rather universal behavior in the transmission coefficients for extremal black holes has also been noticed in the analysis for specific examples of asymptotically AdS solutions, see e.g.  \cite{Denef:2009yy}.

We should emphasize that this sort of phenomenon is not particularly unusual in scattering theory. In fact, there is a familiar setup where these statements can be checked explicitly.  Consider a scattering process governed by the confluent hypergeometric equation. We can then repeat the analysis for this case, except now the connection matrix ${\cal M}_{\infty\to+}$ will just be the identity matrix since there are only two singular points. Still, ${\cal M}_{\rm pw \to +}$ is non-trivial since infinity exhibits Stokes phenomenon.   Then the non-analyticity in \eqref{eq:44} arising from $d_2$ corresponds to multiplication from the right by a diagonal matrix and, therefore, will cancel out when computing the transmission coefficient $|{\cal T}|^2$.  Nevertheless, one can directly verify the branch cuts in the individual elements of the connection matrix by explicitly computing the scattering matrix and seeing that they are, indeed, present.\footnote{A simple example is to study scattering off a Coulomb potential, where this non-analyticity can again be verified.  Finally, it's worth noting that the phenomenon of complex eigenvalues of a Hamiltonian, corresponding to bound states, being confined to a single sheet of the complex energy plane is a very generic feature of quantum mechanical problems; see, e.g., \cite{BohmQM}.}



\section{Future Directions}

In this paper we have developed a procedure, based around monodromy data, for understanding scattering coefficients and quasinormal modes in various black hole backgrounds.
This development allowed us to demonstrate an important result concerning the analytic properties of the scattering coefficients in a Kerr background, also implying exact properties of the equations that govern quasinormal frequencies.  In arriving at this result, we utilized elementary symmetry transformations of the confluent Heun equation. However the ODE has additional integral symmetries which will further restrict the form of the connection matrix \cite{schafke, kazakov};  it would be interesting to exploit these relations in more detail. 

It is rather challenging to obtain analytic properties of the quasinormal mode spectrum. Most of the work on quasinormal modes of rotating non-extremal black holes involves numerical methods \cite{Leaver:1985ax,Nollert:1993zz}, though there are a few analytic methods for the Schwarzschild solution (e.g., \cite{Motl:2002hd,Motl:2003cd,Padmanabhan:2003fx}) and the (near-) extremal Kerr black hole (e.g., see \cite{Glampedakis:2001js,Hod:2008zz} and references within).  Our results, valid in full generality, provide new insights into the complex-analytic properties of these modes. A more detailed comparison with prior work, especially that on extremal black holes, would be intriguing.

In fact, in some sense extremal Kerr could be more tractable than non-extremal: the two horizons merge and form an irregular singular point, yielding a double confluent Heun equation.  The double confluent case has a much larger elementary symmetry group than the confluent case, and an analysis of its implications on scattering data has been nicely summarized in \cite{ronveaux1995}.  Appealing to results from \cite{ronveaux1995} for a scalar probe, one learns that the scattering matrix has two branch cuts: one at $\omega=0$, as for Kerr, and another at $\omega = \frac{m}{2M}$, which is the critical frequency for superradiant scattering and is known to correspond to an accumulation point of quasinormal modes in the extremal limit \cite{Detweiler:1977gy}.  We hope to return to a more detailed discussion of extremal Kerr in future work.

We would like to emphazise that the analytic properties of the ODE governing the perturbations for higher spin modes is in the same universality class as the scalar probe, being described by confluent Heun equations; see appendix \ref{sec:kerrspin}. The technique used for the scalar field in section \ref{sec:exkerr} will then apply for higher spin probes such as gravitational and electromagnetic perturbations. However, the problem of computing scattering coefficients has to be revisited. In particular, the boundary conditions and  reality conditions of the modes  might be modified.  We leave these issues for future work.

Of course, there are many other types of black holes that could be studied using these techniques, including charged black holes, those in AdS or dS spacetimes, and those in higher dimensions.  In many cases, additional singular points appear at complex radii making the analysis more subtle, but in others, such as in AdS, asymptotic infinity is typically regular, so one does not have to worry about Stokes phenomena.  While our exploration of scattering data in black holes focused on neutral, asymptotically flat backgrounds, we hope that the present paper will stimulate further progress in a much wider array of black hole backgrounds.


\section*{Acknowledgments}

{We are grateful to Thomas Faulkner, Gary Gibbons, Sean Hartnoll, Niky Kamran,  Vijay Kumar, Finn Larsen, Don Marolf, Andy Nietzke, Loganayagam Ramalingam, and Reinhard Sch\"afke, for useful conversations. The work of A.M. and A.C. was in part supported by NSF under Grant No. PHY11-25915. AC's work is also supported by the Fundamental Laws Initiative of the Center for the Fundamental Laws of Nature, Harvard University. A.M and J.M.L. are supported by the National Science and Engineering Research Council of Canada. M.J.R. is supported by the European Commission - Marie Curie grant PIOF- GA 2010-275082.}


\appendix


\section{Basic Notation}

Throughout this paper, we discuss ODEs of the form 
\be\label{eq:ODE}
\partial_z \left(\begin{array}{c}\Psi_1\\ \Psi_2\end{array}\right) = \left(\begin{array}{cc}0 & \frac{1}{U(z)} \\ V(z) & 0 \end{array}\right) \left(\begin{array}{c}\Psi_1\\ \Psi_2\end{array}\right) =: A(z) \Psi ~.
\ee
The notation we use to characterize different facets of the system, such as its solutions and global properties, are:
\begin{itemize}
\item $M_{i}$: Monodromy matrix, defined in \eqref{eqn:monodef}.
\item $e^{2\pi i \Lambda_0}$: Formal monodromy matrix, relevant for irregular singular points; see below \eqref{eq:essential-singularity}.
\item $\Phi_{i}$: Fundamental matrix of solutions to \eqref{eq:ODE}, as defined in \eqref{aa:2}. The subscript $i$ could denote that the fundamental matrix diagonalizes $M_i$, as in \eqref{aa:regexpan}, or it could denote one that diagonalizes the formal monodromy $e^{2\pi i \Lambda_0}$ at an irregular singular point, as in \eqref{eqn:FFS}.
\item ${\cal M}_{i\to j}$: Connection matrix between $\Phi_i$ and $\Phi_j$, i.e., $\Phi_i^{-1} \Phi_j$, as in  \eqref{eq:conn-def}.
\item $\alpha_i$: Related to the eigenvalues of the monodromy matrix $M_i$, which are given by $e^{\pm 2\pi \alpha_i}$.  Since ${\rm tr}(M_i) = 2\cosh(2\pi\alpha_i)$ is the well-defined quantity, $\alpha_i$ is only defined up to an overall sign and a shift by an imaginary integer. 
\item $\lambda_i$:  Similar to $\alpha_i$, related to the eigenvalues of the formal monodromy $e^{2\pi i \Lambda_0}$ around an irregular singular point; see either \eqref{eqn:FFS} or  \eqref{app:b2}.
\end{itemize}


\section{Estimating Stokes Multipliers}
\label{app:stokes}

To compute the Stokes multipliers numerically, we use the techniques developed in \cite{daalhuis1995calculation}, which we summarize here.  In the following, we focus on the case of a rank 1 singularity rather than a higher rank singularity, which can be found in \cite{daalhuis1995calculation}.

Consider the differential equation 
\be\label{app:b1}
\partial_z^2 \psi(z)+f(z)\partial_z \psi(z) + g(z) \psi(z) =0~,
\ee
where $f(z)$ and $g(z)$ have convergent series expansions around $z=\infty$.\footnote{In the case that $f(z)$ and $g(z)$ are rational functions, if we call $h(z)$ the least common multiple of their denominators, it is computationally more efficient to multiply \eqref{app:b1} by $h(z)$ since it leads to simpler finite-order recursion relations than otherwise obtained.  This is implemented in the Mathematica notebook we developed \cite{StokesNotebook}.} We will assume that $z=\infty$ is an irregular singular point, so the solutions will have an essential singularity at $z=\infty$. When the rank of the singularity is 1, we can expand $f(z)$ and $g(z)$ as
 \be
 f(z)=\sum_{s=0}^\infty {f_s\over z^s}~,\qquad  g(z)=\sum_{s=0}^\infty {g_s\over z^s}~,
 \ee
where rank equal to 1 implies that at least one of the coefficients $f_0$, $g_0$, or $g_1$, is nonzero.

For simplicity we restrict to the non-resonant case where $f_0^2 \neq 4g_0$, then the two formal series solutions to \eqref{app:b1} are given by
 \be\label{app:b2}
 \psi_1 (z) =e^{\zeta_1 z}z^{ \lambda_1}\sum_{s=0}^\infty {a_{s,1}\over z^s}~,\qquad  \psi_2 (z) =e^{\zeta_2 z}z^{ \lambda_2}\sum_{s=0}^\infty {a_{s,2}\over z^s}~,
 \ee
 where
 \be\label{app:bb3}
 \zeta_{1,2}= -{f_0\over 2}\pm \sqrt{{f_0^2\over 4}-g_0}~.
 \ee
The exponents $\lambda_{1,2}$ are given by
 \be
 \lambda_{1}={f_1\zeta_1+g_1\over \zeta_2-\zeta_1}~,\qquad \lambda_{2}={f_1\zeta_2+g_1\over \zeta_1-\zeta_2}~.
 \ee
 The coefficients of the series expansion \eqref{app:b2} are determined by the recurrence relations
\bea\label{app:b4}
(\zeta_1-\zeta_2)s\,a_{s,1}=(s-\lambda_1)(s-1-\lambda_1)a_{s-1,1}+\sum_{j=1}^{s}\left[\zeta_1f_{j+1}+g_{j+1}-(s-j-\lambda_1)f_j\right]a_{s-j,1}~,\cr
(\zeta_2-\zeta_1)s\,a_{s,2}=(s-\lambda_2)(s-1-\lambda_2)a_{s-1,1}+\sum_{j=1}^{s}\left[\zeta_2f_{j+1}+g_{j+1}-(s-j-\lambda_2)f_j\right]a_{s-j,2}~,
\eea
with the starting values of $a_{0,1}=a_{0,2}=1$ and $a_{s,i} = 0$ for $s<0$.  Generally, $\sum_s a_{s,i} z^{-s}$ will have zero radius of convergence and only corresponds to an asymptotic expansion of a true solution around $z=\infty$. 

In fact, to specify the asymptotic expansion of a true solution around $z=\infty$, we have to specify the argument of $z$ when we take the limit $z\to\infty$ to determine what linear combination of formal solutions $\psi_1$ and $\psi_2$ we will obtain.  Thus, we can think of the formal solutions in (\ref{app:b2}) as corresponding to given true solutions only within a given wedge of the $z$-plane $S_k$, with $k\in\mathbb{Z}$, defined by
\be
S_k=\left\{ z: \left(k-\tfrac{1}{2}\right)\pi-{\rm Arg}\left( \zeta_2-\zeta_1\right) \le {\rm arg}(z) \le \left(k+\tfrac{1}{2}\right)\pi-{\rm Arg}\left( \zeta_2-\zeta_1\right)\right\} \ .
\ee
On a closed sector properly interior to $S_{k-1} \cup S_k \cup S_{k+1}$, there is a single linear relation among the three associated solutions:
\be
\psi_{k+1}(z)=C_k \, \psi_{k}(z)+\psi_{k-1}(z)\ ,
\ee
where the set of constants $\{C_{k}\}$, with $k\in\mathbb{Z}$, are called Stokes multipliers (at infinity) of the differential equation (\ref{app:b1}), and the $\psi_k$ for $k\notin\{1,2\}$ are defined in \eqref{eq:psik}.   Defining $A_k := e^{-(-1)^k k (\lambda_2 - \lambda_1) \pi i} C_k$, we have that
\be
A_k = \left\{ \begin{array}{rcl} A_0 &  & \textrm{for}~ $k$ ~ \textrm{even,}  \\   A_1 & & \textrm{for}~ $k$ ~ \textrm{odd.} \end{array}\right. 
\ee

We also have the relation
\be
\label{eq:psik}
\psi_{k+2j}(z) = \left\{ \begin{array}{rcl} e^{2j \lambda_1 \pi i} \psi_k \big(e^{-2j \pi i}z\big) && \textrm{for $k$ even,} \\
e^{2j \lambda_2 \pi i} \psi_k \big(e^{-2j \pi i}z\big) && \textrm{for $k$ odd.} \end{array} \right.
\ee
Combining these, we see that
\be
\big( \begin{array}{cc} \psi_0(e^{-2\pi i}z) & \psi_1(e^{-2\pi i}z) \end{array} \big) = \big( \begin{array}{cc} \psi_0(z) & \psi_1(z) \end{array} \big) M_\infty~,
\ee
where
\be
M_\infty = \left( \begin{array}{cc} e^{-2\lambda_1 \pi i} & 0 \\ 0 & e^{-2\lambda_2 \pi i} \end{array}\right) \left(\begin{array}{cc} 1 & 0 \\ C_{-1} & 0 \end{array}\right) \left(\begin{array}{cc} 1 & C_0 \\ 0 & 1 \end{array}\right) \ .
\ee
This is just as in \eqref{cc:5}.

We can normalize the independent variable $z$ in such a way that $\zeta_2-\zeta_1=1$ by replacing $z$ with 
\be\label{app:b5}
\frac{z}{\zeta_2-\zeta_1}\,.
\ee 
We implement this in the numerical approach. The values of the Stokes parameters will be related to the original by $(\zeta_2-\zeta_1)^{(-)^{k}(\lambda_2-\lambda_1)} $ times their old value. With the choice \eqref{app:b5}, a good estimate of the Stokes multipliers is
\bea\label{app:b8}
A_0&=&-2\pi i\, a_{s,2}\left[\sum_{j=0}^{m-1}a_{j,1}\Gamma(s-\lambda_2+\lambda_1-j)\right]^{-1} +O (s^{-m})~,\\
A_1&=&2\pi i\, (-1)^{s-1}\,  a_{s,1}\, \left[\sum_{j=0}^{m-1}(-1)^ja_{j,2}\Gamma(s+\lambda_2-\lambda_1-j)\right]^{-1} + O (s^{-m})~.\nonumber
\eea
 For numerical implementation, this is a systematic way to compute $A_{0,1}$.  Alternatively, if the coefficients $a_{s,1}$ and $a_{s,2}$ are known as a function of $s$, the Stokes multipliers can be obtained from the limit
\bea\label{app:b6}
A_0&=&-{2\pi i\over a_{0,1}}\,\lim_{s\to\infty} {a_{s,2}\over \Gamma(s-\lambda_2+\lambda_1)} ~,\cr
A_1&=&{2\pi i\over a_{0,2}} \,\lim_{s\to\infty} (-1)^{s-1}{a_{s,1}\over \Gamma(s+\lambda_2-\lambda_1)}~.
\eea

When using the summation formula \eqref{app:b8}, which is a very efficient way to compute the Stokes multipliers, it's worth noting that the large $s$ behavior of the coefficients $a_{s,i}$ implied by \eqref{app:b6} suggests a restriction on the choice of $m$ in \eqref{app:b8} not mentioned in \cite{daalhuis1995calculation}.  The sum over $j$ in \eqref{app:b8} would clearly diverge for $m=\infty$, so one should truncate the sum around the smallest term.  If we use the asymptotic behavior of $a_{s,i}$ to get a rough idea where to truncate, this suggests choosing
\be
m \,  \lesssim  \, \frac{s+1}{2} \mp {\rm Re}(\lambda_2 - \lambda_1) \, ,
\ee
where the upper sign is for $A_0$ and the lower for $A_1$.  This is only a rough bound, so it is good to check different values of $m$ and $s$ to verify the consistency of any given computation.  Obviously, this also implies a rough bound on the choice of $s$, namely
\be
s \gtrsim {\rm max}\big\{1 , \, 1 \pm 2 {\rm Re}(\lambda_2 - \lambda_1) \big\} \ .
\ee

\subsection{Numerical Implementation for Schwarzschild and Kerr}\label{app:numKS}

The above algorithm can be implemented in a straightforward manner for the ODEs relevant to Schwarzschild and Kerr backgrounds; the Mathematica code can be found in  \cite{StokesNotebook}. For concreteness, we computed the monodromy data \eqref{cc:omega},  \eqref{R2}, for certain values of quasinormal frequencies.  In Tables \ref{tab:myfirsttable} and \ref{tab:mysecondtable}, we used the values of $\omega_\qnm$ reported in \cite{Berti:2009kk}. 

\begin{table}[t]
\centering
\scalebox{0.8}{
\begin{tabular}{ |c|c|c|c|c|c|}
\hline   $n$ & $\ell$ & $\omega^{{\rm Sch}}_\qnm$ &${\cal T}_1{\cal T}_2$&${\cal R}_1{\cal R}_2$& $\alpha_\irr$\\
\hline  0 & 0 & $0.220910-0.209791 \, i$ &$0.725516 - 1.3456\,i$&$0.981555 + 0.183844\,i$&$0.10688-0.00489413 \,i$ \\
 \hline 1 & 0 & $0.172234-0.696105 \, i$ &$1.99733 - 0.327733\,i$&$0.703811 + 0.213504\,i$&$0.184325 - 0.395024\, i$\\  
 \hline 2 & 0 & $0.151484-1.20016 \,i$ &$1.90982 - 0.0988799\,i$&$0.638641 + 0.171458 \,i$&$0.196496 + 0.139823\,i$\\  
 \hline
\end{tabular}
}
\caption{This table shows the QNM of the Schwarzschild black hole ($a=0$) of mass $M=\frac{1}{2}$ where $n$ is the overtone and $\ell$ is the total angular momentum quantum number.}
\label{tab:myfirsttable}
\end{table}

\begin{table}[t]
\centering
\scalebox{0.615}{
\begin{tabular}{ |c|c|c|c|c|c|c| }
\hline    $a/M$& $\ell = m$ & $\omega^{{\rm Kerr}}_\qnm$&$K_\ell$&${\cal T}_1{\cal T}_2$&${\cal R}_1{\cal R}_2$&$\alpha_\irr$ \\
\hline   0.2 & 0  &$0.221535 - 0.209025 i$&$-0.0000179424 + 0.00030871 i$&$0.720005 - 1.34114 i$&$0.979579 + 0.17628 i$&$0.104914 - 0.0124874 i$\\
 \hline  0.4 & 0 &$0.223398 - 0.206506 i$  &$-0.0000966287 + 0.00123024 i$&$0.703135 - 1.32729 i$&$0.97394 + 0.152931 i$&$0.106939 - 0.0019500 i$\\ 
 \hline  0.6 & 0 & $0.226342 - 0.201397 i$ &$-0.000319103 + 0.00273531 i$&$0.673668 - 1.3024i$&$0.966199 + 0.11148 i$&$0.106438 + 0.0022231 i$\\
 \hline  0.8 & 0 &$0.229074 - 0.191402 i$ &$-0.000841989 + 0.0046779 i$&$0.628351 - 1.2633  i$&$0.964207 + 0.0471877 i$&$0.10359 + 0.00897525 i$\\
 \hline  0.96 & 0  &$0.222904 - 0.178774 i$&$-0.001356610 + 0.0061231 i$&$0.576185 - 1.22289 i$&$0.997748 - 0.00361162 i$&$0.094805 + 0.0129196 i$\\ 
 \hline  0.98 & 0  &$0.221232 - 0.178961 i$&$-0.00134879 + 0.00633965 i$&$0.571018 - 1.21868 i$&$1.00052 - 0.00027528 i$&$0.0940499 + 0.0121104 i$\\ 
  \hline  0.99 & 0  &$0.220894 - 0.178998 i$&$-0.00136306 + 0.00646113 i$&$0.568935 - 1.21602 i$&$0.999991 + 0.0000324433 i$&$0.0938947 + 0.0119539 i$\\ 
  \hline  0.999 & 0  &$0.220768 - 0.178795 i$&$-0.00138927 + 0.00656798 i$&$ 0.569172 - 1.2139 i$&$ 1. + 2.44249 \times10^{-15} i$&$0.0937403 + 0.0120083 i$\\   
  \hline  0.9999 & 0  &$0.218527 - 0.181398 i$&$-0.00123152 + 0.00660759  i$&$0.54043 - 1.21994 i$&$1. + 1.63758 \times10^{-15} i$&$0.09376 + 0.00974293 i$\\ 
    \hline  0.9999 & 2(n=1) &$ 1.98647 - 0.00705074 i$&$5.85717 + 0.00102783 i$&$ 1.00493 - 0.000359751i$&$7.06779\times10^{-6} + 0.000264006  i$&$0.944334 + 0.498461 i$\\ 
  \hline  0.9999 & 2(n=2)  &$1.98644 - 0.021193 i$&$5.85719 + 0.00308939 i$&$ 1.00503 - 0.0107268i$&$0.000207279 + 0.000347074 i$&$0.946068 + 0.495206 i$\\   
  \hline  0.9999 & 2(n=3)  &$1.98635 - 0.0353136 i$&$5.85724 + 0.00514751  i$&$0.994687 - 0.0176253 i$&$0.000379994 + 0.000312636 i$&$0.949419 + 0.491461 i$\\   \hline
\end{tabular}
}
\caption{This table shows the QNM of the Kerr black hole, with mass $M=\frac{1}{2}$, for different values of the black hole's angular momentum $\frac{a}{M}\in[0,1]$ and values of $\ell=m$. The overtone is $n=1$ unless stated otherwise.}
\label{tab:mysecondtable}
\end{table}

\begin{figure}[t]
\centering{
\includegraphics{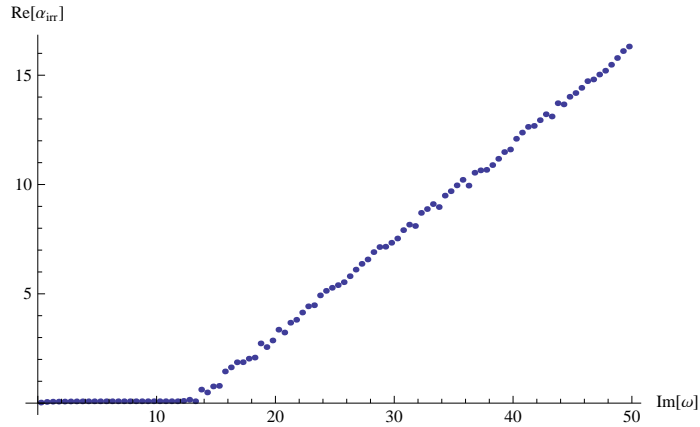}}
\caption{The figure shows the real part of the monodromy around the irregular singular point ${\rm Re}(\alpha_\irr)$ as a function of the highly damped QNM frequency ${\rm Im}( \omega_\qnm)=\pi \, T_{BH} (2n-1)$ for fixed values of the mass $M=\frac{1}{2}$, angular momenta $a=0$, and $\ell=1$. The {\it blue dots} are the numerical values;  for $n\gg 40$, the growth becomes linear and ${\rm Re}(\alpha_{irr}) \approx 0.43243\,{\rm Im}(\omega_\qnm)-5.51604$.}
\label{fig:alphaschw}
\end{figure}

In addition to the low-lying quasinormal modes, we can also implemented the numerics for highly damped frequencies.  Some interesting results on QNM for Schwarzschild black holes \cite{Motl:2002hd} showed that in the high overtone limit $n\rightarrow\infty$, the frequencies are approximately given by $\omega_\qnm \approx T_{BH}\ln(3)+i \pi \, T_{BH} (2n-1)$, where $T_{BH}=(8\pi M)^{-1}$ is the temperature.  In this regime, we find that for large values of the eigenvalue $\ell$, the monodromy $\alpha_\irr$ satisfies
\be
{\rm Re}(\alpha_\irr)=a\,{\rm Im}(\omega_\qnm)+b,~\qquad a,b \in \mathbb{R}~.
\ee
The plot in Fig. \ref{fig:alphaschw} illustrates this linear relation.

\section{Perturbative Monodromy at Irregular Singular Points}
\label{sec:perturbation}

Suppose we have an ODE
\be
\big(\p_z - A(z)\big) \Phi(z) = 0~,
\ee
with an irregular singular point at $z=\infty$, and suppose we can split the connection in two as
\be\label{eq:splitA}
A(z) = A^{(0)}(z) + \epsilon A^{(1)}(z) \, ,
\ee
where $\epsilon$ is a small parameter and $A^{(0)}(z)$ is a connection whose exact solutions are known, which we group into a fundamental matrix $\Phi^{(0)}(z)$.  Then we can write $\Phi(z) = \Phi^{(0)}(z) \Phi^{(1)}(z)$, where $\Phi^{(1)}(z)$ satisfies the ODE
\be
\big(\p_z - \epsilon A_{\epsilon} \big) \Phi^{(1)}(z) := \big(\p_z - \epsilon\Phi^{(0)}{}^{-1} A^{(1)} \Phi^{(0)} \big) \Phi^{(1)}(z) = 0 \ .
\ee
$A_\epsilon$ is what we would call the `interaction-picture Hamiltonian' if we were talking about the Schr\"odinger equation (where $A$ would be anti-Hermitian and $\Phi^{(0)}$ would be unitary).

We want to compute the conjugacy class of the monodromy around $z=\infty$, which is uniquely described by its trace and determinant. By construction, the determinant is one while the trace is given by 
\be
{\rm tr} M_\infty =  {\rm tr}\Big( \Phi(z)^{-1} \Phi(z e^{-2\pi i})\Big) =  {\rm tr}\Big( \Phi^{(1)}(z)^{-1} \Phi^{(0)}(z)^{-1} \Phi^{(0)}(z e^{-2\pi i}) \Phi^{(1)}(z e^{2\pi i})\Big) \ ,
\ee
where we take $z\to z e^{-2\pi i} $ to circle $z=\infty$ in the positive direction.  Since we know exactly what $\Phi^{(0)}(z)$ is, we can exactly compute the constant matrix
\be
M^{(0)}_\infty := \Phi^{(0)}(z)^{-1} \Phi^{(0)}(z e^{-2\pi i}) \ .
\ee
Thus, we have
\be\label{eq:trace}
{\rm tr}M_\infty =  {\rm tr}\Big( M_\infty^{(0)} \Phi^{(1)}(z e^{-2\pi i}) \Phi^{(1)}(z)^{-1} \Big) =  {\rm tr}\Big( M_\infty^{(0)} {\cal P} \big\{ e^{\epsilon \oint_\gamma A_\epsilon} \big\} \Big) \ .
\ee
We can then expand the exponential and obtain ${\rm tr}M_\infty$ as a series expansion in $\epsilon$.  In fact, the right-hand side should be independent of a choice of $\gamma$ within a given homotopy class, so we can take $\gamma$ to be a circle centered on $z=\infty$ and take the limit as $z\to \infty$ to render the computations tractable.

As we will see in both examples below, this method is very well suited for low-frequency expansions. It would be interesting to adapt a similar perturbative technique to high frequency regimes; in particular, it could be instructive to compare our technique with those  used in \cite{Motl:2002hd,Motl:2003cd,Musiri:2003bv,Musiri:2007zz} to compute highly damped quasinormal mode frequencies.  

\subsection{Modified Angular Equation: $a\omega\ll1$ limit}
\label{sec:angeqn}

Introducing a coordinate $z :=\cos\theta$, we can rewrite the angular equation \eqref{bb:angeqn} as
\be
\label{eq:modified-angular2}
\p_z\big((1-z^2)\p_z{S}(z)\big) + \Big( K_\ell + (a\omega)^2 z^2 - \frac{m^2}{1-z^2}\Big) {S}(z) \, ,
\ee
which has an irregular singular point at $z=\infty$ and two regular singular points at $z=\pm 1$ with monodromy eigenvalues $(-1)^{m}$.

Before we begin computing the monodromy around $z=\infty$, let's consider the other singularities first and understand the problem we are trying to solve.  The monodromies around the regular singular points, $z=\pm 1$, are conjugacy equivalent to
\be
M_{\pm 1} \cong \smat (-1)^m & 1 \\ 0 & (-1)^m \esmat \ .
\ee
Both solutions have a square-root branch cut when $m$ is odd, but the nontrivial Jordan block implies more, it implies that one solution also has a logarithmic branch cut.  We are looking for solutions that are regular at \emph{both} $z=1$ and $z=-1$, which correspond to $\theta=0,\pi$.  Since the regular solution at $z=1$ ($z=-1$) corresponds to the only eigenvector of $M_{1}$ ($M_{-1}$), this means that $K_\ell$ must take values that allow for $M_1$ and $M_{-1}$ to share the same eigenvector.  Recalling that we can write $M_{\pm 1}$ in a common basis as (see \eqref{eq:cbcb})%
\be
 M_{-1}=\left(\begin{array}{cc}0&-1\\ 1&2(-1)^m\end{array}\right) ~,\qquad\quad M_{+1} = \left(\begin{array}{cc}2(-1)^m & e^{2\pi  \alpha_\irr}\\ -e^{-2\pi  \alpha_\irr}&0\end{array}\right)~,  
\ee
we see they share an eigenvector when
\be\label{eq:trecond}
e^{2\pi\alpha_\irr} = 1  \quad \Longrightarrow \quad  \alpha_\irr \in i\bb{Z} \quad {\rm and} \quad {\rm tr}M_\infty = 2 \ .
\ee
Computing the monodromy at $z=\infty$ will then let us determine the values of $K_\ell$ that satisfy this condition.

To compute $M_\infty$, we will use \eqref{eq:trace}. 
We can write \eqref{eq:modified-angular2} as a linear ODE with connection
\bea
A(z) &= &\left(\begin{array}{cc} 0 & \frac{1}{z^2-1}  \\  k_{0} + \frac{m^2}{z^2-1} & 0  \end{array}\right) dz  +  (a\omega)^2 \left(\begin{array}{cc} 0 & 0 \\ z^2 + \delta K_\ell & 0 \end{array}\right) dz\cr &&\cr&  =:& A^{(0)}(z) + (a\omega)^2 A^{(1)}(z) \ ,
\eea
where we split $K_\ell$ as
\be
K_\ell= k_0 +(a\omega)^2 \delta K_\ell~, \qquad k_0 = \ell(\ell+1) \ .
\ee
The fundamental matrix for $A^{(0)}$ can be written as
\be
\Phi^{(0)}(z) = \left( \begin{array}{cc}  S_1(z)  &  S_2(z)  \\  (z^2-1)\p_z S_1(z)  &  (z^2-1)\p_z S_2(z) \end{array}\right) \ ,
\ee
where
\bea
S_1(z) &=&  z^{\delta+m-\frac{1}{2}} (z^2-1)^{-\frac{m}{2}} \ {}_2 F_1(\alpha,\beta,1-\delta; z^{-2})  \nonumber \\
S_2(z) &=& z^{-\delta+m-\frac{1}{2}} (z^2-1)^{-\frac{m}{2}} \ {}_2 F_1(\alpha+\delta,\beta+\delta,1+\delta; z^{-2})
\eea
and
\be
\delta = \tfrac{1}{2}\sqrt{1+4k_0} \, , \qquad \alpha=\tfrac{1}{4}-\tfrac{m}{2}-\tfrac{1}{2}\delta \, ,  \qquad  \beta = \tfrac{1}{2}+\alpha \ .
\ee
In this case, then,
\be
M_\infty^{(0)} = \left(\begin{array}{cc} -e^{-\pi i\sqrt{1+4k_0}} & 0 \\ 0 & -e^{\pi i\sqrt{1+4k_0}} \end{array}\right) \ .
\ee
To obtain ${\rm tr} \, M_\infty$, we can now expand the right-hand side of (\ref{eq:trace}) as a series in $\epsilon := (a\omega)^2$ so that
\be
{\rm tr} \, M_\infty = {\rm tr} \, M_\infty^{(0)} + (a\omega)^2 \oint dz \, {\rm tr}\big\{ M^{(0)}_\infty A_\epsilon(z) \big\} + O\big((a\omega)^4\big) \ ,
\ee
where $A_\epsilon(z) :=\Phi(z)^{-1}A^{(1)}(z)\Phi(z)$.  Calling $\textrm{tr}M_\infty = 2\cosh(2\pi\alpha_\irr)$, this leads to
\bea
2\cosh(2\pi\alpha_\irr) &=& -2\cos(\pi\sqrt{1+4k_0}) \cr &&- a^2\omega^2 2\pi\sin(\pi\sqrt{1+4k_0}) \tfrac{3+4m^2+8\delta K_\ell - (1+4k_0)(1+2\delta K_\ell )}{(4k_0-3)\sqrt{1+4k_0}} + O\big((a\omega)^4\big)~. \qquad
\eea
Using $k_0 = \ell(\ell+1)$, we have
\be
\alpha_\irr= i\ell + i a^2\omega^2 \, \tfrac{2\ell(\ell+1)(1+2\delta K_\ell)-2m^2-1-3\delta K_\ell}{(2\ell+1)(4\ell(\ell+1)-3)} + O\big((a\omega^4)\big) \ .
\ee

As we saw, to have a solution that is regular at both poles, we must set $\alpha_\irr\in i\Z$, which determines $\ell \in \Z$ and 
\bea
\delta K_\ell = \frac{2m^2+1-2\ell(\ell+1)}{(2\ell-1)(2\ell+3)} + O(a^2\omega^2) \ .
\eea
So, as we expect, $K_\ell = \sum_{n\geq 0} k_n (a\omega)^{2n}$ where
\be
k_0 = \ell(\ell+1) \ ,  \qquad  k_1 = \frac{2m^2+1-2\ell(\ell+1)}{(2\ell-1)(2\ell+3)} \ .
\ee
The restriction of $-\ell \leq m \leq \ell$ comes from a more careful treatment of normalizability of the respective eigenfunctions than we gave here.  One can continue this to higher order straightforwardly.  For use in the next subsection, we quote the next coefficient from \cite{Abramowitz},
\be
k_2 = \tfrac{(\ell-m-1)(\ell-m)(\ell+m-1)(\ell+m)}{2(2\ell-3)(2\ell-1)^3(2\ell+1)}-\tfrac{(\ell-m+1)(\ell-m+2)(\ell+m+1)(\ell+m+2)}{2(2\ell+1)(2\ell+3)^3(2\ell+5)} \ .
\ee
It is also worth mentioning that Mathematica has a built in function for computing $K_\ell(m,a\omega)$, called {\it SpheroidalEigenvalue}.


\subsection{Modified Radial Equation: $M\omega\ll1$ limit}
\label{app:monKerr}
\newcommand{\veps}{\varepsilon}

Recall the radial part of the scalar Kerr wave equation from \eqref{bb:10}:
\bea
\left[
\partial_r \Delta \partial_r
+{({2Mr_+\omega}-{a}m )^2\over (r-r_+)(r_+-r_-)}   - {({2Mr_-\omega}-{a}m )^2\over (r-r_-)(r_+-r_-)}
 + (r^2+2M(r+2M) )\omega^2 \right] R(r)=K_\ell  R(r)~ . \nonumber
\eea
Next, define
\be
  a:=M \sqrt{1-\veps^2}  ~~ \Longrightarrow ~~ r_\pm = M(1\pm\varepsilon) \ ,  \qquad  0\leq\veps\leq 1 \ ,
\ee
so $\veps=1$ is Schwarzschild and $\veps=0$ is extremal Kerr (where the two horizons merge and become an irregular singular point).
In terms of these parameters, the horizon monodromies are
\be
\label{eq:coordxform}
\alpha_\pm := {{2Mr_\pm\omega}-{a}m \over r_+-r_-} = \frac{2M\omega(1\pm \veps)-m \sqrt{1-\veps^2}}{2\veps} ~.
\ee

Next, for the non-extremal case, we define\footnote{For the computations in this section, we found it slightly more convenient to choose coordinates where the irregular singular point is at $z=0$ rather than at infinity.  We then call its monodromy matrix $M_\irr$ to emphasize that it is the monodromy at the irregular singular point.  This is what we called $M_\infty$ in other sections.}
\be
r=:\frac{2M\veps}{z} + r_- \ ,  \qquad \hat{R}(z) := z^{-1/2}R\big(r(z)\big) \ ,
\ee
so that the equation becomes
\bea\label{eq:kerr-che}
\bigg[\p_z z(1-z)\p_z + \frac{\alpha_+^2}{1-z} -\frac{1+4\alpha_-^2}{4} - \frac{1+4K_\ell-4M^2\omega^2(7-4\veps+\veps^2)}{4z}  \qquad\qquad &&  \\
+ \frac{4M^2\omega^2\veps^2}{z^3}+\frac{4\veps M^2\omega^2(2-\veps)}{z^2} \bigg]\hat{R}(z) &=& 0 \ , \nonumber
\eea
with the inner horizon at $z=\infty$, the outer horizon at $z=1$, and ``asymptotic infinity'', the irregular singular point, at $z=0$.  We can turn this into a first order ODE with connection
\be\label{eq:splitK}
\hat{A}(z) = \left(\!\begin{array}{cc} 0 & \frac{1}{z(1-z)}      \\   \frac{1+4\alpha_-^2}{4} - \frac{\alpha_+^2}{1-z} + \frac{1+4K_\ell-(2M\omega)^2(7-4\veps+\veps^2)}{4z}  & 0  \end{array}\!\right)
  +  (2M\omega)^2 \left(\! \begin{array}{cc} 0 & 0  \\   \frac{\veps(\veps-2)}{z^2} -\frac{\veps^2}{z^3}  & 0 \end{array} \!\right) \, . 
\ee
Writing this as $\hat{A}(z) = \hat{A}^{(0)}(z) + (2M\omega)^2 \hat{A}^{(1)}(z)$, we choose $\hat{A}^{(0)}(z)$ to be the first term in \eqref{eq:splitK}. The solutions to the ODE with connection $A^{(0)}(z)$ are
\be
\hat{\Phi}^{(0)}(z) = \left(\begin{array}{cc} \hat{R}_1(z)  &  \hat{R}_2(z)  \\  z(1-z)\p_z\hat{R}_1(z)  &  z(1-z)\p_z \hat{R}_2(z) \end{array}\right) \ ,
\ee
where 
\bea
\hat{R}_1(z) &=& z^{\frac{\gamma}{2}} \ (1-z)^{\frac{\alpha+\beta-\gamma-1}{2}}\ {}_2 F_1(\alpha,\beta;1+\gamma;z) =:  z^{\frac{\gamma}{2}} \ (1-z)^{\frac{\alpha+\beta-\gamma-1}{2}}\ F_1(z) \ , \\
\hat{R}_2(z) &=& z^{-\frac{\gamma}{2}} \ (1-z)^{\frac{\alpha+\beta-\gamma-1}{2}} \ {}_2 F_1(\alpha-\gamma,\beta-\gamma;1-\gamma;z) =:  z^{-\frac{\gamma}{2}} \ (1-z)^{\frac{\alpha+\beta-\gamma-1}{2}} \ F_2(z) \ .  \nonumber
\eea
This satisfies
\be
\det\hat{\Phi}^{(0)}(z) = -\gamma
\ee
for all $z$.  The constants appearing as arguments of the hypergeometric function are given by
\begin{gather}
\alpha = \tfrac{1+\gamma}{2} + i(\alpha_+ + \alpha_-) ~,  \qquad  \beta = \tfrac{1+\gamma}{2} + i(\alpha_+ - \alpha_-) ~ , \nonumber \\[0.2cm]
\gamma = \sqrt{1+4K_\ell-(2M\omega)^2(7-4\veps+\veps^2)} ~ .
\end{gather}
Similar to the angular equation, because the transformation $R = \sqrt{z}\hat{R}$ has branch cuts, the trace of the monodromy for the original system \eqref{bb:10} differs from that of $\hat{A}(z)$ by an overall minus sign.  Thus,
\be
M_\irr^{(0)} = - \hat{\Phi}^{(0)}(z)^{-1} \hat{\Phi}^{(0)}(z e^{2\pi i}) =  -\left(\begin{array}{cc} (-1)^{\gamma} & 0 \\ 0 & (-1)^{-\gamma}  \end{array}\right) = -e^{i\pi\gamma\sigma^3} \ .
\ee

Now we want to begin computing the Dyson series for the connection
\be
A_\epsilon(z) :=   \Phi^{(0)}(z)^{-1} \left(\begin{array}{cc} 0 & 0 \\  -\frac{\veps^2}{z^3}-\frac{\veps(2-\veps)}{z^2}   & 0  \end{array}\right) \Phi^{(0)}(z) \ ,
\ee
where we identify the $\epsilon$ expansion parameter (not to be confused with $\varepsilon$) with $(2M\omega)^2$.  Writing this in a slightly more useful way,
\be
A_\epsilon(z) :=  \frac{\veps(\veps+(2-\veps) z)}{\gamma \, z^3} \ (1-z)^{2i\alpha_+} \ z^{-\frac{\gamma}{2}\sigma^3}  \left(\begin{array}{cc} -F_1(z) F_2(z) & -F_2(z)^2 \\  F_1(z)^2   &  F_1(z)F_2(z)  \end{array}\right) z^{\frac{\gamma}{2}\sigma^3} ~ .
\ee
We must keep in mind the validity of the order to which we work: each term in the Dyson series comes with a power of $(2M\omega)^2$.  On the other hand, $M$ and $\omega$ appear in the solution $\Phi^{(0)}$ as well, so the answer is only meaningful after expanding these constants to the appropriate power of $\omega$.
Finally,
\bea
{\rm tr} \, M_\irr &=& -2\cos\pi\gamma + (2M\omega)^2 \oint dz \, {\rm tr}\big( M^{(0)}_\irr A_\Omega(z) \big)  \nonumber \\
&& \quad + \tfrac{1}{2} (2M\omega)^4 \oint dz_2 \oint dz_1 \, {\rm tr}\Big( M^{(0)}_\irr\, \mathcal{P}\big\{A_\Omega(z_2) A_\Omega(z_1)\big\} \Big) + O(\omega^5) \ .
\eea
Keeping terms of order no more than $\omega^4$, we find
\be
\label{eqn:monodromy}
\alpha_\irr^2 ~:=~ \Big(\tfrac{1}{2\pi}\cosh^{-1}\big(\tfrac{1}{2}{\rm tr}M_\irr\big)\Big)^2 
~=~ -\ell^2 +\sum_{n=1} a_n (2M \omega)^n \ , 
\ee
where the first few coefficients for $\ell \neq 0$ are
\bea
a_1&=&0~,\cr
a_2&=& \ell \frac{15\ell(\ell+1)-11}{(2\ell+3)(2\ell+1)(2\ell-1)}~, \cr
a_3&=&  - 2 m  \sqrt{1-\veps^2}\ \frac{5\ell(\ell+1)-3}{(\ell+1)(2\ell+3)(2\ell+1)(2\ell-1)}~, \cr
a_4&=&\frac{1}{4(\ell+1)(2\ell+5)(2\ell+3)^3(2\ell+1)^3(2\ell-1)^3(2\ell-3)} \Big( 80 (\ell(\ell+1))^5 \big[115+26 \veps^2\big] \nonumber \\
&& \qquad -8 (\ell(\ell+1))^4 \big[6971+450 \ell+1234 \veps^2-1020 m^2 \big(1-\veps^2\big)\big] \nonumber \\
&& \qquad +(\ell(\ell+1))^3 \big[97075+21480 \ell+8058 \veps^2-38928 m^2 \big(1-\veps^2\big)\big] \nonumber \\
&&\qquad -(\ell(\ell+1))^2 \big[35821 \ell+6 \big(11827+60 \veps^2-5529 m^2 \big(1-\veps^2\big)\big)\big]  \nonumber \\
&& \qquad +3 \ell(\ell+1) \big[7387+7854 \ell-252 \veps^2-702 m^2 \big(1-\veps^2\big)\big] \nonumber \\
&& \qquad -45 \big[49+121 \ell + 72 m^2 \big(1-\veps^2\big)\big] \Big) \, .
\eea
The leading term when $\ell=m=0$ is instead given by
\be
\alpha_\irr^2 = -\tfrac{49}{36}(2M\omega)^4 + O(\omega^5)  \ .
\ee
The expression for the monodromy is also valid in the limit $\veps\to1$, which corresponds to the Schwarzschild solution, though the intermediate steps require a slight modification because the parameters of the hypergeometric functions $\alpha$ and $\beta$ are equal, requiring special treatment.  The limit $\veps\to 0$ may require a separate treatment since the coordinate transformation \eqref{eq:coordxform} is singular; although, the low-frequency expansion we obtain above certainly has a well-defined limit $\veps \to 0$, so it may also give the correct expression for the monodromy in the extremal case.

Another approach to compute $\alpha_\irr$ was discussed in \cite{Mano:1996vt}, where they reported only the leading correction in $\omega$. Their discussion involved studying the recursion relations for a Floquet solution of the type \eqref{bb:basetrue}. A careful analysis of the low-frequency limit allows one to use their results to solve for $\alpha_\irr$ up to $O(\omega^2)$, which agree with the values of $a_1$ and $a_2$ computed above.


\section{Schwarzschild  Black Hole}\label{app:SBH}

In Schwarzschild coordinates we have
\be\label{app:schwgeom}
ds^2=- \left(1+{2M\over r}\right)dt^2+\left( 1-{2M\over r}\right)^{-1}dr^2+ r^2d\theta^2+r^2\sin^2\theta d\phi^2~.
\ee
The Klein--Gordon equation for a massless scalar  is again \eqref{bb:KG}, and
expanding in eigenmodes
\be
\psi(t,r,\theta, \phi) = e^{-i\omega t+im\phi}R(r)S(\theta)~,
\ee
the spherical function $S(\theta)$ satisfies
\be
\left[ {1\over \sin\theta}\partial_\theta
\left(\sin\theta\,\partial_\theta\right)-{m^2\over\sin^2\theta}\right] S(\theta)=-K_\ell  S(\theta)~,
\ee
whose solution is the associated Legendre polynomial $P^m_l(\cos\theta)$ with eigenvalue $K_\ell=\ell(\ell+1)$. The radial equation for $R(r)$ is
\bea
\left[
\partial_r \,r\,(r-2M) \partial_r
+{(2M)^3{\omega}^2\over (r-2M)}  
 + (r^2+2M(r+2M) )\omega^2 \right] R(r)=K_\ell  R(r)~ .
\eea
The singular points of the ODE for Schwarzschild corresponds to the coordinate singularity  $r=0$, the  horizon $r=2M$, and infinity $r\to\infty$.

To determine the scattering coefficients, we first must find the constants $\alpha_+$, $\alpha_-$, and $\alpha_\irr$, which are logarithms of the eigenvalues of the monodromy matrices associated with each of the singular points (see equations (\ref{eq:matrices1}) and (\ref{eq:matrices2})). We can find this information by studying 
the behavior of the solutions $R(r)$ near these points.
Around the regular singular points $r=0$ and $r=2M$, the behavior of the solutions is respectively
\bea
&&R(r) = (r-2M)^{ i \alpha_+}(a_1+\ldots)+  (r-2M)^{- i \alpha_+}(a_2+\ldots)~,\qquad  \alpha_+= {2M\omega} ~,\nonumber\\
&&R(r) = r^{ i \alpha_-}(b_1+ b_3 \log r +\cdots)+  r^{- i \alpha_-}(b_2+\cdots) ~,\qquad  \alpha_-= 0~.
\eea
Note that the singularity at $r=0$ is a resonant regular singular point (hence the logarithm).
The singularity at $r\to \infty$ is an irregular singular point of rank 1, and the problem of finding $\alpha_\irr$ will be analogous to that of Kerr, so it can be determined by (\ref{cc:omega}). 
While the product of the Stokes multipliers $C_0C_{-1}$ will differ, the formal monodromy will be the same as that for Kerr, $\lambda=2M\omega$.

\section{Higher Spin Perturbations}
\label{sec:kerrspin}

Define eigenmodes for a spin-$s$ field $\psi_s$ as
\be\label{eq:spin}
\psi_s(t,r,\theta,\phi)= e^{i\omega t+m\phi} R_s(r) S_s(\theta) \ ,
\ee
which are related to scalar ($s=0$), electromagnetic ($s=\pm1$), and gravitational ($s=\pm 2$) perturbations around the Kerr background. The detailed relation between $\psi_s$ and the field perturbations can be found in  \cite{Staronbinski:1974,Teukolsky:1974yv}. 

For $s=0,\pm 1,\pm 2$, we define $R_s(r) := \Delta^{-s/2}\hat{R}_s(r)$. Then the radial part of the wave equation for a particle of spin $s$ can be written as  
\bea
&\!\!\!\!\!\!\!\!\!\!&\Bigg[\p_r \Delta \p_r +  \frac{\left(2M\omega r_+ - \frac{i}{2} s(r_+ - r_-) - a m\right)^2}{(r-r_+)(r_+ - r_-)}  - \frac{\left(2M\omega r_- + \frac{i}{2} s (r_+ - r_-) - a m\right)^2}{(r-r_-)(r_+ - r_-)}    \qquad\qquad\qquad\qquad  \\
&\!\!\!\!\!\!\!\!\!\!& \quad\qquad\qquad\qquad\qquad\qquad + \omega^2 r^2 + 2(M\omega+is)\omega r + 2M\omega(2M\omega - is) - s^2 - K_{\ell,s} \Bigg] \hat{R}_s(r) = 0 \ .  \nonumber
\eea 
Again, $K_{\ell,s}$ is the separation constant, i.e., it corresponds to an eigenvalue of the spin-weighted spheroidal equation
\be\nonumber
\left[\frac{1}{\sin\theta} \p_\theta \left( \sin\theta \p_\theta \right) + (a\omega\cos\theta-s)^2 - \frac{(m+s\cos\theta)^2}{\sin^2\theta} - s^2 + K_{\ell,s} \right] S_s(\theta) = 0  \ .
\ee
Notice that the radial ODE is not real when $s\neq 0$; instead, complex conjugation (for real parameters and real $r$) is only a symmetry if coupled with $(a,m,s) \to (-a,-m,-s)$. 

The form of the radial equation is the same as before: a confluent Heun equation with regular singular points at the two horizons and an irregular singular point of rank 1 at infinity, so the same techniques from the spin 0 analysis carry over here.  The monodromies of $\hat{R}$ around the horizons are given by
\be\label{eq:mspin}
\hat{\alpha}_\pm = \mp \frac{is}{2} + \frac{2M\omega  - a m}{r_+ - r_-} \ .
\ee
Around infinity,
\be
\hat{R}(r) \sim e^{\pm i \omega r} r^{\pm i(2 M\omega + is) - 1} \ ,
\ee
so in the notation of section \ref{sec:Heun-symmetries}, we have
\be
\hat{\varpi} = \omega(r_+ - r_-) \ ,  \qquad  \hat{\lambda} = 2 M \omega + is \ ,  \qquad  \hat{\kappa} = K_{\ell,s} + s^2 + (a^2-8M^2)\omega^2 \ .
\ee
Notice that $\hat{\alpha}_+ \pm \hat{\alpha}_-$ differ from their $s=0$ values by an imaginary integer, since $s\in\mathbb{Z}$, and similarly for $\hat{\lambda}$, so reality properties of quantities such as $\sinh\!\pi(\hat{\alpha}_+ \pm \hat{\alpha}_-)$ are unaffected.  

Massless fermionic perturbations ($s=\pm1/2$) need a separate treatment \cite{Chandrasekhar:1976ap, Page:1976jj}. Furthermore, reality conditions  and boundary conditions are sightly different in this case. However, an expression like \eqref{eq:mspin} still applies.


\bibliographystyle{utphys}
\bibliography{monodromies-bib}

\end{document}